\documentclass[10pt, twocolumn, notitlepage, superscriptaddress, prb, longbibliography]{revtex4-2}

\usepackage{here}
\usepackage{amsmath}         
\usepackage{graphicx}
\usepackage{braket}         
\usepackage{lettrine}
\usepackage{xcolor}
\usepackage{times}
\usepackage{hhline}
\usepackage{tabularx}
\usepackage{calc}
\usepackage{soul}
\usepackage{float}

\usepackage[colorlinks=true,linkcolor=blue, citecolor=blue, urlcolor=blue]{hyperref}%

\usepackage[T1]{fontenc}

\begin{document}

\title{Orbital Pumping by Magnetization Dynamics in Ferromagnets}    

\author{Dongwook Go}
\email{dongo@uni-mainz.de}
\affiliation{Institute of Physics, Johannes Gutenberg University Mainz, 55099 Mainz, Germany}

\author{Kazuya Ando}
\affiliation{Department of Applied Physics and Physico-Informatics, Keio University, Yokohama 223-8522, Japan}
\affiliation{Keio Institute of Pure and Applied Sciences (KiPAS), Keio University, Yokohama 223-8522, Japan}
\affiliation{Center for Spintronics Research Network (CSRN), Keio University, Yokohama 223-8522, Japan}

\author{Armando Pezo}
\affiliation{Laboratoire Albert Fert, CNRS, Thales, Université Paris-Saclay, Palaiseau 91767, France}

\author{Stefan Bl\"ugel}
\affiliation{Peter Gr\"unberg Institut, Forschungszentrum J\"ulich, 52425 J\"ulich, Germany}

\author{Aur\'elien Manchon}
\affiliation{Aix-Marseille Université, CNRS, CINaM, Marseille, France}

\author{Yuriy Mokrousov}
\affiliation{Institute of Physics, Johannes Gutenberg University Mainz, 55099 Mainz, Germany}
\affiliation{Peter Gr\"unberg Institut, Forschungszentrum J\"ulich, 52425 J\"ulich, Germany}

\begin{abstract}
We show that dynamics of the magnetization in ferromagnets can pump the orbital angular momentum, which we denote by orbital pumping. This is the reciprocal phenomenon to the orbital torque that induces magnetization dynamics by the orbital angular momentum in non-equilibrium. The orbital pumping is analogous to the spin pumping established in spintronics but requires the spin-orbit coupling for the orbital angular momentum to interact with the magnetization. We develop a formalism that describes the generation of the orbital angular momentum by magnetization dynamics within the adiabatic perturbation theory. Based on this, we perform first-principles calculation of the orbital pumping in prototypical $3d$ ferromagnets, Fe, Co, and Ni. The results show that the ratio between the orbital pumping and the spin pumping ranges from 5 to 15 percents, being smallest in Fe and largest in Ni. This implies that ferromagnetic Ni is a good candidate for measuring the orbital pumping. Implications of our results on experiments are also discussed.
\end{abstract}

\date{\today}                 
\maketitle	      

Injection of spin into a ferromagnet can induce magnetization dynamics, which is called the spin torque~\cite{Slonczewski1996, Berger1996, Myers1999, Stiles2002, Ralph2008}. The reciprocal effect known as the spin pumping can generate non-equilibrium spin by magnetization dynamics~\cite{Tserkovnyak2002a, Tserkovnyak2002b, Mizukami2002, Azevedo2005, Saitoh2006}. The spin torque and pumping have played pivotal roles in spintronics as means of manipulating the magnetization and generating spin currents, respectively~\cite{Tserkovnyak2005, Brataas2017}. In general, however, the physical mechanism of magnetization dynamics is governed by transfer of the \emph{angular momentum}~\cite{Haney2010, Go2020b}, which is not restricted to the spin. As a result, a recent theory has shown that non-equilibrium orbital angular momentum (OAM) can also induce magnetization dynamics~\cite{Go2020a}, which is denoted by the orbital torque. A crucial difference between the spin torque and the orbital torque is that the orbital torque requires the spin-orbit coupling (SOC) because local moments in a ferromagnet are coupled to spin by the exchange interaction, and the OAM interacts with magnetization \emph{indirectly} via  SOC. Thus, the orbital torque depends sensitively on the way the spin and orbital characters are entangled for the states near the Fermi energy~\cite{Go2020b, Lee2021a}. The orbital torque has been measured in several experiments in recent years, via the sign change, crucial dependence on the choice of a ferromagnet, and orbital-to-spin conversion~\cite{Ding2020, Kim2021, Lee2021a, Lee2021b, Sala2022, Dutta2022, Liao2022, Hayashi2023a, Bose2023}.

\begin{figure}[b!]
\centering
\vspace{-11pt}
\includegraphics[angle=0, width=0.4\textwidth]{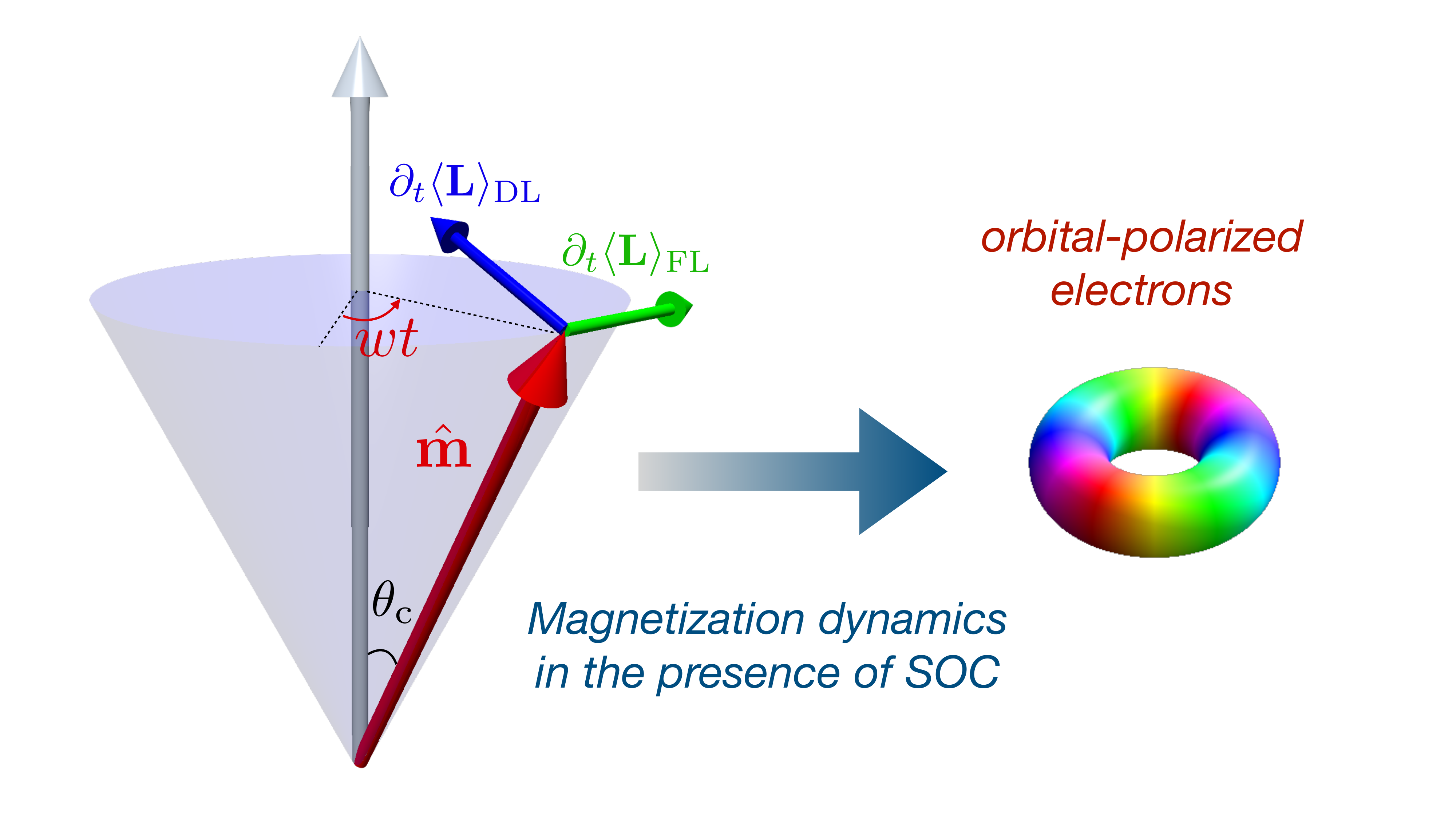}
\caption{
\label{fig:orbital_pumping_schematics}
Concept of the orbital pumping.
}
\end{figure}

A natural question that arises is what are the properties of the reciprocal phenomenon to the orbital torque. We may denote this by \emph{orbital pumping}, generation of non-equilibrium OAM from dynamics of the magnetization, whose concept is schematically illustrated in Fig.~\ref{fig:orbital_pumping_schematics}. Recently, Hayashi \emph{et al.} reported an experimental observation of the orbital pumping in Ti/Ni bilayers~\cite{Hayashi2023b}. Here, the ferromagnetic resonance generates the OAM together with the spin in Ni, which is injected across the interface. The injected OAM in Ti is electrically measured by using the inverse orbital Hall effect, where the inverse spin Hall effect is negligibly small due to small SOC~\cite{Salemi2022, Choi2023, Go2024}. This result is consistent with the recent observation of large orbital Hall effect in Ti~\cite{Choi2023} and the orbital torque measurement which finds pronounced torque response when Ni is used as a ferromanget~\cite{Hayashi2023a}. Interestingly, Hayashi \emph{et al.} found that the signal is significantly suppressed in Ti/Fe bilayers~\cite{Hayashi2023b}, implying that the orbital pumping is less pronounced in Fe. One may ask what makes the difference of the orbital pumping between Fe and Ni. This motivates us to investigate its microscopic origin and perform quantitative evaluation of the orbital pumping in $3d$ ferromagnets.

In this Letter, we develop a formalism describing the orbital pumping within the adiabatic perturbation theory and derive a Green's function expression for the generation of the OAM by magnetization dynamics. From first-principles, we compute the orbital pumping in ferromagnetic Fe, Co, and Ni. We show that the orbital pumping is a concomitant effect of the spin pumping due to the SOC and thus exhibits a Hund-rule-like behavior~\cite{Hund1925, Levine2014}, evolving from the negative to the positive value as the occupation number of the $d$ shell increases. The magnitude of the orbital pumping ranges from 5 to 15 percents of that of the spin pumping in Fe, Co, and Ni, being smallest in Fe and largest in Ni. This result explains the recent experiment measuring significant orbital pumping contribution in Ti/Ni~\cite{Hayashi2023b}, but nearly complete suppression of the signal in Ti/Fe cannot be explained by the bulk property of Fe alone.

We define the orbital pumping as the response of $\partial_t \langle \mathbf{L} \rangle$ to the magnetization dynamics, where $\mathbf{L}$ is the OAM operator. This can be formally written as a linear response expression,
\begin{eqnarray}
\partial_t 
\left \langle L_\alpha \right \rangle 
=
\sum_\beta
\chi_{\alpha\beta}^L
\left( 
\hat{\mathbf{m}} \times \partial_t \hat{\mathbf{m}}
\right)_\beta,
\label{eq:orbital_pumping}
\end{eqnarray}
where $\hat{\mathbf{m}}$ is the unit vector of the magnetization, and $\alpha,\beta$ are Cartesian indices. Note that $\hat{\mathbf{m}}\times \partial_t \hat{\mathbf{m}}$ acts as a generalized force, pointing in the direction of the Gilbert damping. As shown in Fig.~\ref{fig:orbital_pumping_schematics}, $\partial_t \langle \mathbf{L} \rangle$ can be conveniently decomposed into fieldlike and dampinglike components, which point in the directions of $\partial_t \hat{\mathbf{m}}$ and $\hat{\mathbf{m}}\times \partial_t \hat{\mathbf{m}}$, respectively. Let us consider a situation of the magnetization precessing around a certain axis (grey arrow) with the cone angle $\theta_\mathrm{c}$ and angular frequency $w$. For convenience, let us define  Cartesian coordinates such that $x$, $y$, and $z$ are axes parallel to $\partial_t \hat{\mathbf{m}}$, $\hat{\mathbf{m}}\times \partial_t \hat{\mathbf{m}}$, and $\hat{\mathbf{m}}$, respectively. From Eq.~\eqref{eq:orbital_pumping}, we find $\partial \langle \mathbf{L} \rangle_\mathrm{FL}= \hat{\mathbf{x}}\chi_{xy}^L w\sin\theta_\mathrm{c}$ and $\partial \langle \mathbf{L} \rangle_\mathrm{DL}=\hat{\mathbf{y}}\chi_{yy}^L w\sin\theta_\mathrm{c}$, where $w\sin\theta_\mathrm{c}$ is the norm of $\hat{\mathbf{m}}\times \partial_t \hat{\mathbf{m}}$. Both the fieldlike and dampinglike components have AC responses, but the projection of the dampinglike component on the precession axis generates a DC response, which is given by $\chi_{yy}^L w\sin^2\theta_\mathrm{c}$. We emphasize that assuming that $\theta_\mathrm{c}$ is small enough, $\chi_{\alpha\beta}^L$ may be computed for a fixed configuration of $\hat{\mathbf{m}}$ along an easy-axis. 


To evaluate $\chi_{\alpha\beta}^L$ from the electronic structure, we apply the adiabatic perturbation theory~\cite{Vanderbilt2018}, in which $\hat{\mathbf{m}}$ evolves slowly enough for the electronic eigenstates to evolve adiabatically. From this, we derive a Green's function's expression for $\chi_{\alpha\beta}^L$~\cite{Supplemental} in the Bastin-Smr\v{c}ka-St\v{r}eda's form~\cite{Crepieux2001,Bastin1971,Smrcka1977} of the Kubo formula~\cite{Kubo1956},
\begin{subequations}
\begin{eqnarray}
\label{eq:Green_function_expression_a}
\chi_{\alpha\beta,\mathrm{even}}^L
&=&
\frac{\hbar}{4\pi}
\int d\mathcal{E} \ \left( \partial_\mathcal{E} f \right)
\\
&\times&
\mathrm{Tr}
\left[ 
T_\mathrm{tot}^{L_\alpha}
\left(
G^\mathrm{R}  - G^\mathrm{A} 
\right)
T_\mathrm{XC}^{S_\beta}
\left(
G^\mathrm{R}  - G^\mathrm{A} 
\right)
\right],
\nonumber 
\\
\label{eq:Green_function_expression_b}
\chi_{\alpha\beta,\mathrm{odd}}^L
&=&
\frac{\hbar}{4\pi}
\int d\mathcal{E} f 
\\
&\times&
\mathrm{Tr}
\left[ 
T_\mathrm{tot}^{L_\alpha}
\left( G^\mathrm{R}  - G^\mathrm{A} \right) 
T_\mathrm{XC}^{S_\beta}
\left(
\partial_\mathcal{E} G^\mathrm{R}  + \partial_\mathcal{E} G^\mathrm{A} 
\right)
\right.
\nonumber
\\
& &
\ \ \ \ 
\left.
-
T_\mathrm{tot}^{L_\alpha}
\left( 
\partial_\mathcal{E} G^\mathrm{R}  + \partial_\mathcal{E} G^\mathrm{A} 
\right)
T_\mathrm{XC}^{S_\beta}
\left(
G^\mathrm{R}  - G^\mathrm{A} 
\right)
\right],
\nonumber
\end{eqnarray}
\label{eq:Green_function_expression}
\end{subequations}
where $G^\mathrm{R/A} =1/(\mathcal{E}-\mathcal{H}_\mathrm{tot}\pm i\Gamma)$ is the retarded/advanced Green's function, $\mathcal{H}_\mathrm{tot}$ is the total Hamiltonian, $\mathcal{E}$ is the energy. The energy broadening $\Gamma$ effectively describes relaxation of the electron quasiparticles by scatterings. The Fermi-Dirac distribution function $f$ is defined at energy $\mathcal{E}$. The operators $T_\mathrm{tot}^{L_\alpha}=[L_\alpha,\mathcal{H}_\mathrm{tot}]/i\hbar$ and $T_\mathrm{XC}^{S_\beta}=[S_\beta, \mathcal{H}_\mathrm{XC}]/i\hbar=J(\mathbf{r})\hat{\mathbf{m}}\times\mathbf{S}$ are the total torque on the $\alpha$ component of the OAM and the \emph{exchange} torque on the $\beta$ component of the spin, respectively. Here, $\mathcal{H}_\mathrm{XC} = J(\mathbf{r})\ \hat{\mathbf{m}}\cdot\mathbf{S}$ is the exchange potential. In Eq.~\eqref{eq:Green_function_expression}, we adopt the decomposition depending on the time-reversal parity of the response~\cite{Bonbien2020}. Meanwhile, the analogous expression of Eq.~\eqref{eq:orbital_pumping} for the spin pumping can be obtained by replacing $L_\alpha$ by $S_\alpha$ in Eq.~\eqref{eq:Green_function_expression}, which agrees with the expression derived in Refs.~\cite{Simanek2003, Mills2003}.

On the other hand, the linear response expression for the orbital torque is given by
\begin{eqnarray}
\left\langle T_\mathrm{XC}^{S_\alpha} \right\rangle 
=
\sum_\beta \widetilde{\chi}_{\alpha\beta}^L
B_\beta^{L},
\end{eqnarray}
where $B_\beta^L$ is an effective field that couples to the OAM by the Zeeman interaction $\mathcal{H}_L =  \mathbf{B}_L \cdot \mathbf{L}$. It effectively describes a situation where non-equilibrium OAM is induced, for example, by orbital Hall effect~\cite{Bernevig2005, Kontani2009, Go2018} or orbital Edelstein effect~\cite{Yoda2018, Salemi2019, Johansson2021}. Note that both $\partial_t \langle \mathcal{H}_L \rangle =\partial_t \langle \mathbf{L} \rangle \cdot \mathbf{B}^L$ and $\partial_t \langle \mathcal{H}_\mathrm{XC} \rangle = \langle T_\mathrm{XC}^\mathbf{S} \rangle \cdot (\hat{\mathbf{m}}\times\partial_t{\hat{\mathbf{m}}}) = \langle J \mathbf{S} \rangle \cdot \partial_t \hat{\mathbf{m}}$ have the unit of \emph{power}. The expression for $\widetilde{\chi}_{\beta\alpha}^L$ can also be derived similarly to Eq.~\eqref{eq:Green_function_expression}, by which the reciprocal relation 
\begin{eqnarray}
\chi_{\alpha\beta}^L = \widetilde{\chi}_{\beta\alpha}^L
\end{eqnarray}
can be explicitly shown~\cite{Supplemental}.

\newcolumntype{P}[1]{>{\centering\arraybackslash}p{#1}}
\newcolumntype{M}[1]{>{\centering\arraybackslash}m{#1}}
\renewcommand{\arraystretch}{1.3}
\begin{table}[b!]
\begin{center}
\centering
\begin{tabular}{M{2.5cm} | M{2.5cm} M{2.5cm}}
\hline \hline
Material & $\chi_{xx}^L/\chi_{xx}^S\ [\%]$ & $\chi_{xy}^L/\chi_{xy}^S\ [\%]$ \\
\hline
Fe (25 meV) & $6.08$ & $4.38$ \\
Co (25 meV) & $9.87$ & $9.62$ \\
Ni (25 meV) & $11.08$ & $15.10$ \\
\hline
Fe (100 meV) & $7.19$ & $4.39$ \\
Co (100 meV) & $9.53$ & $9.32$ \\
Ni (100 meV) & $10.64$ & $14.12$ \\
\hline \hline
\end{tabular}
\caption{
\label{tab:ratio}
The ratio between the orbital pumping and spin pumping in Fe, Co, and Ni for the diagonal and off-diagoanl responses,  obtained for $\Gamma = 25\ \mathrm{meV}$ and $100\ \mathrm{meV}$ of the energy broadening. 
}
\end{center}
\end{table}

\begin{figure*}[t!]
\centering
\includegraphics[angle=0, width=0.8\textwidth]{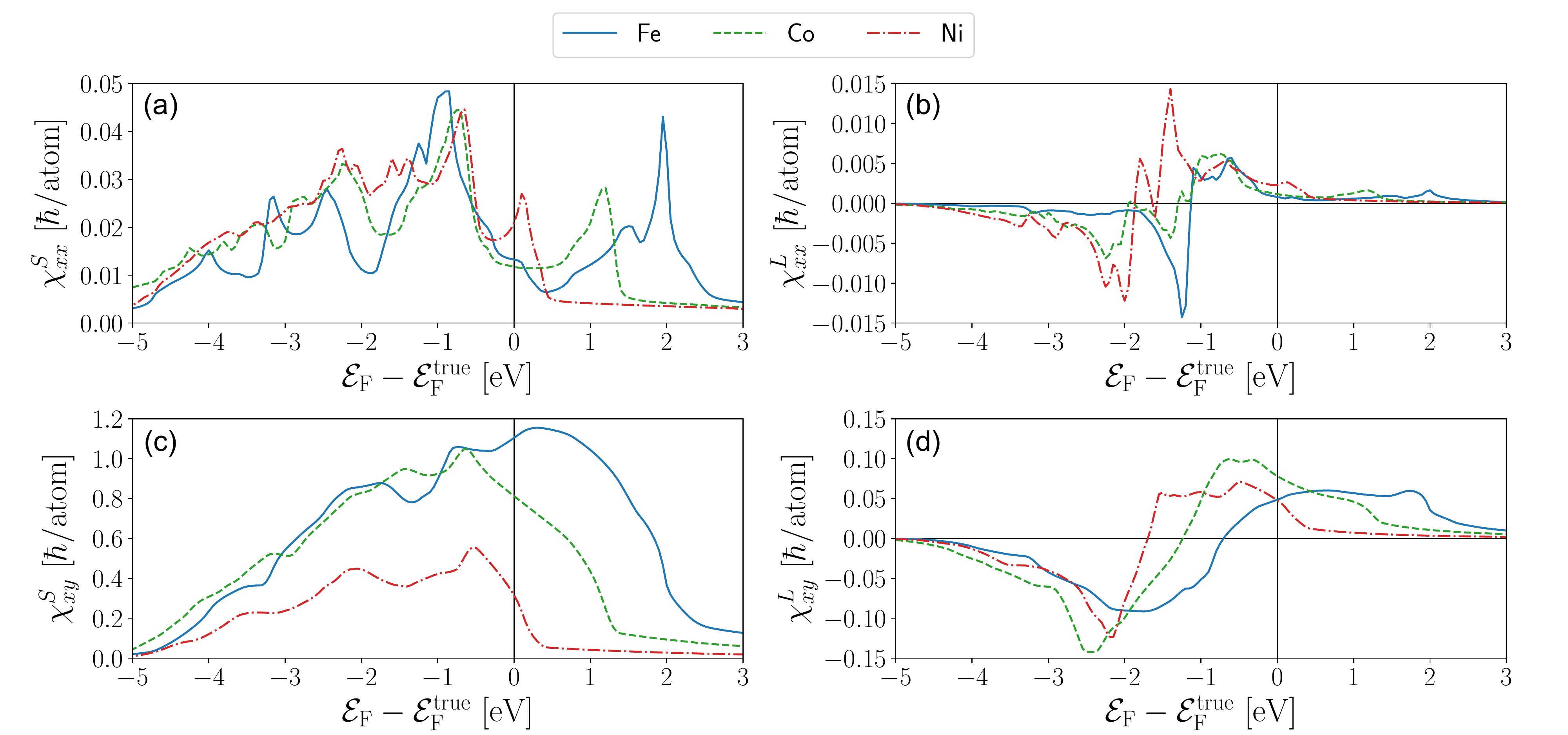}
\caption{
\label{fig:Fermi_dep}
The plots of the Fermi energy ($\mathcal{E}_\mathrm{F}$) dependence of (a) $\chi_{xx}^S$ and (b) $\chi_{xx}^L$, for the spin pumping and orbital pumping, respectively. The same plots for $\chi_{xy}^S$ and $\chi_{xy}^L$ are shown in (c) and (d), respectively. The energy broadening is set to $\Gamma=25\ \mathrm{meV}$ in evaluating Eq.~\eqref{eq:Green_function_expression}. The Fermi energy is varied with respect to the true value $\mathcal{E}_\mathrm{F}^\mathrm{true}$ within the rigid band approximation. 
}
\end{figure*}

We quantitatively evaluate the spin pumping and orbital pumping in prototypical $3d$ ferromagnets: bcc Fe, hcp Co, and fcc Ni. We set the magnetization $\hat{\mathbf{m}}=-\hat{\mathbf{z}}$. Details of the computational method can be found in Ref.~\cite{Supplemental}. For all Fe, Co, and Ni, only $\chi^L_{xx,\mathrm{even}}$ and $\chi^L_{yy,\mathrm{even}}$ are nonzero for the $\mathcal{T}$-even part [Eq.~\eqref{eq:Green_function_expression_a}], and only $\chi^L_{xy,\mathrm{odd}}$ and $\chi^L_{yx,\mathrm{odd}}$ are nonzero for the $\mathcal{T}$-odd part [Eq.~\eqref{eq:Green_function_expression_b}], where $\mathcal{T}$ is the time-reversal. This constraint comes from the $\mathcal{T}\mathcal{M}$ symmetry of the systems, where $\mathcal{M}$ is the reflection with respect to a mirror plane containing the magnetization~\cite{Supplemental}. Therefore, we drop ``even'' and ``odd'' in the subscript of $\chi^L_{\alpha\beta}$ and $\chi^S_{\alpha\beta}$ in the discussion below.

The results for the ratio $\chi_{xx}^{L}/\chi_{xx}^{S}$ and $\chi_{xy}^{L}/\chi_{xy}^{S}$ are summarized in Tab.~\ref{tab:ratio}. The ratios are between 5 and 15 percents, and the magnitudes are in the order of $\text{Fe}<\text{Co}<\text{Ni}$. We find that the ratios are nearly the same for the two chosen values of the energy broadening $\Gamma$, $25\ \mathrm{meV}$ and $100\ \mathrm{meV}$, implying that the relative magnitude between the orbital pumping and spin pumping is robust with respect to the degree of disorder. In order to understand what makes Fe, Co, and Ni different, we investigate Fermi energy dependence of both diagonal ($\chi_{xx}^S$ and $\chi_{xx}^L$) and off-diagonal ($\chi_{xy}^S$ and $\chi_{xy}^L$) components of the response, which are shown in Figs.~\ref{fig:Fermi_dep}(a,b) and Figs.~\ref{fig:Fermi_dep}(c,d), respectively. The sign for the spin pumping [Figs.~\ref{fig:Fermi_dep}(a,c)] remains positive over a wide range of energy because the spin pumping is mainly due to the exchange spin splitting. The orbital pumping, however, exhibits gradual change of the sign, from negative to positive values [Figs.~\ref{fig:Fermi_dep}(b,d)]. We confirm that the orbital pumping is linearly proportional to the SOC strength and vanishes if the SOC is switched off~\cite{Supplemental}. That is, the orbital pumping is a secondary effect following the spin pumping by the SOC. This also explains the small magnitude of the orbital pumping [Figs.~\ref{fig:Fermi_dep}(b,d)] compared to the the spin pumping [Figs.~\ref{fig:Fermi_dep}(a,c)] by an order of magnitude. 

\begin{figure}[b!]
\centering
\includegraphics[angle=0, width=0.4\textwidth]{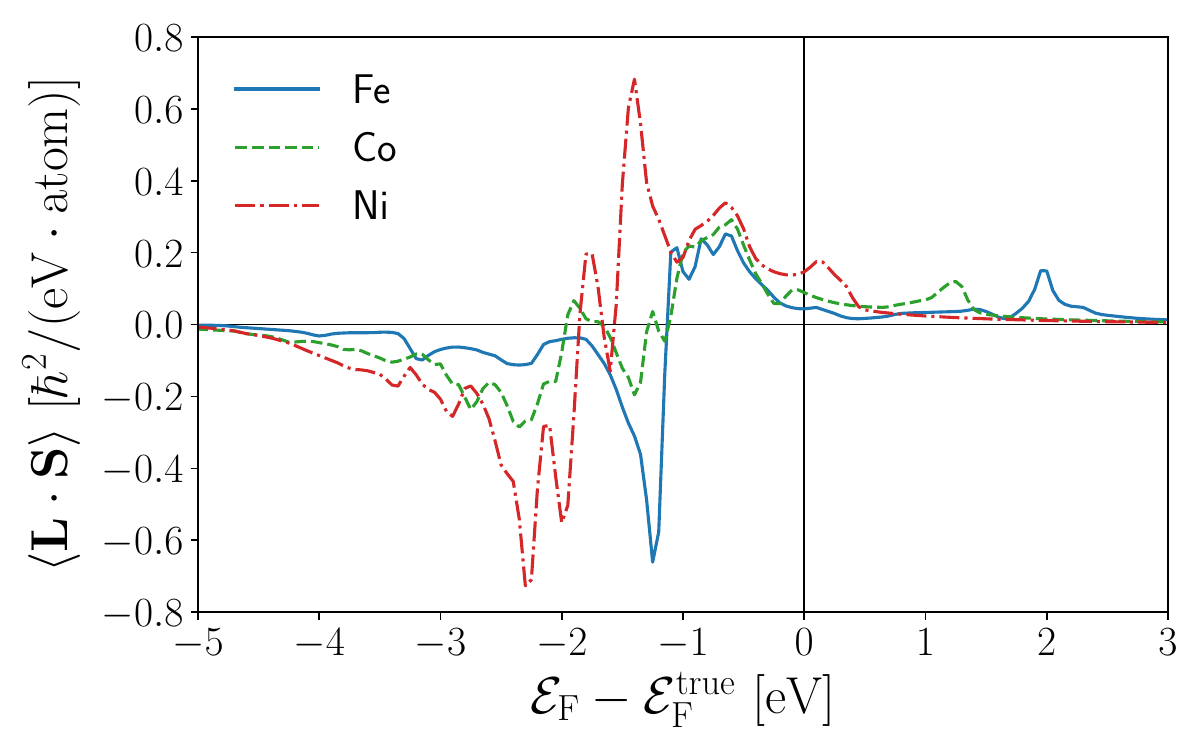}
\caption{
\label{fig:LS_correl}
Correlation between $\mathbf{L}$ and $\mathbf{S}$ at the Fermi surface.
}
\end{figure}

For transition metals, Hund's third rule~\cite{Hund1925, Levine2014} states that when the $d$ shell is approximately less than half-filled, the correlation between $\mathbf{L}$ and $\mathbf{S}$ is negative, $\langle \mathbf{L}\cdot\mathbf{S} \rangle <0$, thus the spin pumping and orbital pumping exhibit the opposite signs. When the $d$ shell is more than half-filled, the correlation becomes positive $\langle \mathbf{L}\cdot\mathbf{S} \rangle >0$, thus the spin pumping and orbital pumping have the same sign. The correlations between $\mathbf{L}$ and $\mathbf{S}$ for the states at the Fermi surface in Fe, Co, and Ni are shown in Fig.~\ref{fig:LS_correl}. In all Fe, Co, and Ni, $\langle \mathbf{L}\cdot\mathbf{S} \rangle$ evolve from negative values to positive values as the Fermi energy gradually increases. Note the stark resemblance between Fig.~\ref{fig:Fermi_dep}(b) and Fig.~\ref{fig:LS_correl}, which clearly demonstrates the crucial role of the correlation between the orbital and spin angular momenta in the orbital pumping.

In general, we find that the diagonal component [Figs.~\ref{fig:Fermi_dep}(a,b)] exhibits more rapid variation as a function of the energy, as compared to the off-diagonal component [Figs.~\ref{fig:Fermi_dep}(c,d)]. This is because the $\mathcal{T}$-even~[Eq.~\eqref{eq:Green_function_expression_a}] and $\mathcal{T}$-odd response~[Eq.~\eqref{eq:Green_function_expression_b}] capture the contributions at the Fermi \emph{surface} and from the Fermi \emph{sea}, respectively. At zero temperature, Eq.~\eqref{eq:Green_function_expression_a} becomes
\begin{eqnarray}
\label{eq:analytic_even}
\chi_{\alpha\beta,\mathrm{even}}^L
&=&
\frac{\hbar}{\pi} \sum_\mathbf{k}\sum_{nm} A_{n\mathbf{k}}(\mathcal{E}_\mathrm{F}) A_{m\mathbf{k}}(\mathcal{E}_\mathrm{F})
\\
& &
\times
\mathrm{Re}
\left[
\bra{\psi_{n\mathbf{k}}} T_\mathrm{tot}^{L_\alpha} \ket{\psi_{m\mathbf{k}}}
\bra{\psi_{m\mathbf{k}}} T_\mathrm{XC}^{S_\beta} \ket{\psi_{n\mathbf{k}}}
\right],
\nonumber
\end{eqnarray}
where $\psi_{n\mathbf{k}}$ is the Bloch state with band index $n$ and crystal momentum $\mathbf{k}$, and $A_{n\mathbf{k}}(\mathcal{E}_\mathrm{F})=\Gamma/[(\mathcal{E}_\mathrm{F}-\mathcal{E}_{n\mathbf{k}})^2+\Gamma^2]$ is the spectral function at the Fermi energy, and $\mathcal{E}_{n\mathbf{k}}$ is the energy dispersion. On the other hand, the $\mathcal{T}$-odd response [Eq.~\eqref{eq:Green_function_expression_b}] has \emph{intrinsic limit} even in the zeroth order of $\Gamma$,
\begin{eqnarray}
\label{eq:analytic_odd}
\chi_{\alpha\beta,\mathrm{odd}}^L
&\approx &
{\hbar}\sum_\mathbf{k}\sum_{n\neq m} 
(f_{n\mathbf{k}}-f_{m\mathbf{k}})
\\
& & 
\times 
\frac{
\mathrm{Im}
\left[
\bra{\psi_{n\mathbf{k}}} T_\mathrm{tot}^{L_\alpha} \ket{\psi_{m\mathbf{k}}}
\bra{\psi_{m\mathbf{k}}} T_\mathrm{XC}^{S_\beta} \ket{\psi_{n\mathbf{k}}}
\right]
}{(\mathcal{E}_{n\mathbf{k}}-\mathcal{E}_{m\mathbf{k}})^2},
\nonumber
\end{eqnarray}
where $f_{n\mathbf{k}}$ is the Fermi-Dirac distribution function.

\begin{figure}[t!]
\centering
\includegraphics[angle=0, width=0.45\textwidth]{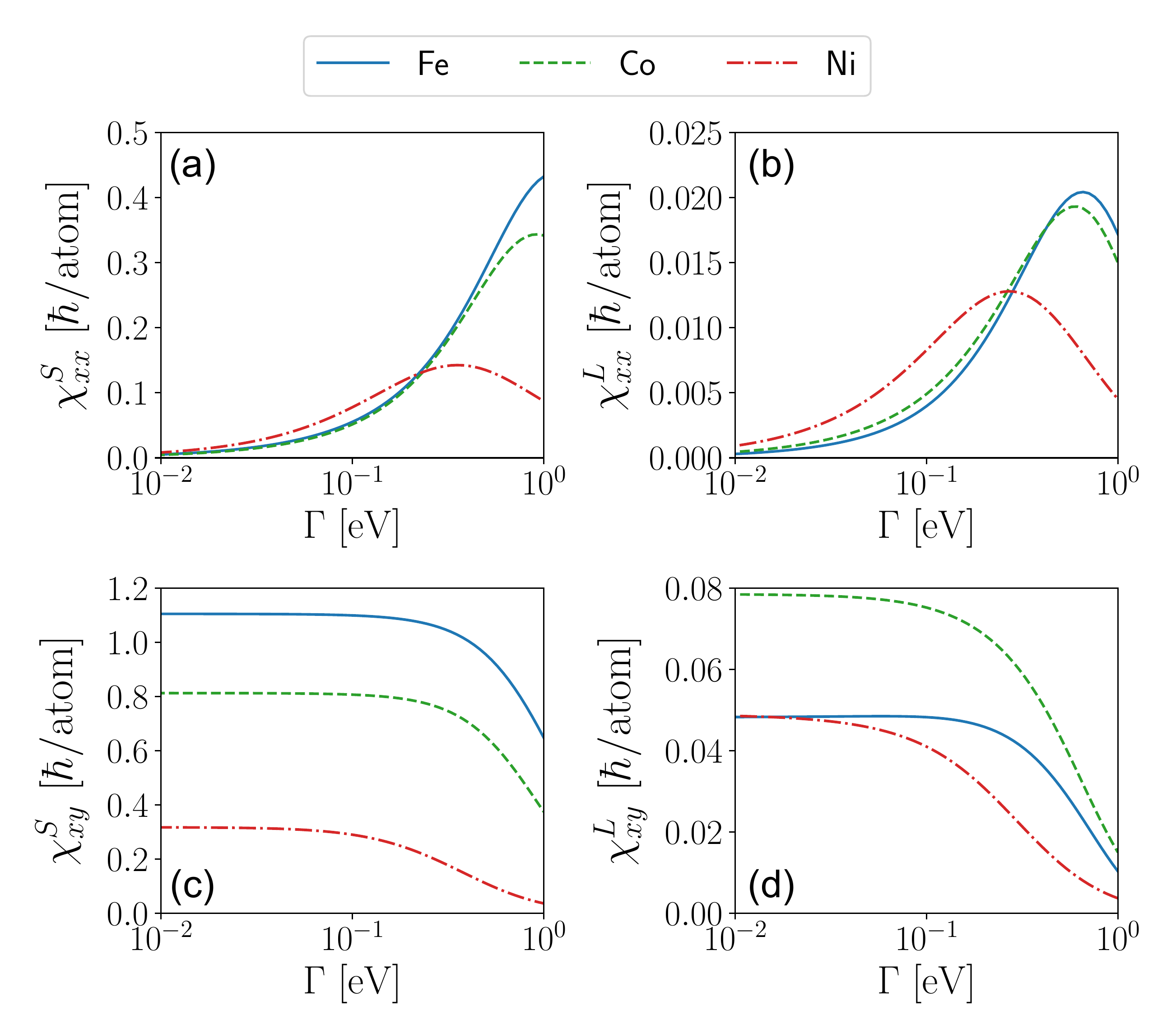}
\caption{
\label{fig:Gamma_dep}
The plots of the energy broadening ($\Gamma$) dependence of (a) $\chi_{xx}^S$ and (b) $\chi_{xx}^L$, for the spin pumping and orbital pumping, respectively. The same plots for $\chi_{xy}^S$ and $\chi_{xy}^L$ are shown in (c) and (d), respectively.
}
\end{figure}

Because of this, the diagonal and off-diagonal responses exhibit different behaviors as a function of the energy broadening $\Gamma$. Figures~\ref{fig:Gamma_dep}(a) and \ref{fig:Gamma_dep}(b) show the $\Gamma$-dependence of $\chi_{xx}^S$ and $\chi_{xx}^L$, respectively. Interestingly, both of them exhibit non-monotonic behavior. They tend to increase from the ultraclean regime ($\Gamma \approx 0$) to moderately clean regime ($\Gamma\sim 10^{-1}\ \mathrm{eV}$), but as $\Gamma$ further increases toward a disordered regime ($\Gamma\sim 1\ \mathrm{eV}$) the magnitudes start to decrease. This suggests that samples need to be moderately clean but not too clean to have sizable magnitude of $\chi_{xx}^S$ and $\chi_{xx}^L$. We emphasize that expressions like Eq.~\eqref{eq:analytic_even} often exhibit a conductivity-like $\propto 1/\Gamma$ behavior because the band-diagonal term ($n=m$) is proportional to $A_{n\mathbf{k}}(\mathcal{E}_\mathrm{F}) A_{n\mathbf{k}}(\mathcal{E}_\mathrm{F})\approx (\pi/\Gamma) \delta (\mathcal{E}_\mathrm{F}-\mathcal{E}_{n\mathbf{k}})$ in small $\Gamma$ limit~\cite{Freimuth2014, Bonbien2020}. However, $\chi_{xx}^S$ and $\chi_{xx}^L$ show a quadratic $\propto \Gamma^2$ behavior because the band-diagonal component of the torque-torque correlation in Eq.~\eqref{eq:analytic_even} is absent. Therefore, spectral mixing of bands at the Fermi energy is crucial for $\chi_{xx}^S$ and $\chi_{xx}^L$. We note that non-monotonic dependence on the broadening has been found in the study of the Gilbert damping~\cite{Hankiewicz2007, Gilmore2007}.

On the other hand, $\chi_{xy}^S$ and $\chi_{xy}^L$ shown in Figs.~\ref{fig:Gamma_dep}(c) and \ref{fig:Gamma_dep}(d), respectively, are stable up to moderate strength of the broadening, displaying a typical intrinsic behavior. In general, we find that the spin pumping [Figs.~\ref{fig:Gamma_dep}(a,c)] is more robust with respect to $\Gamma$ when compared to the orbital pumping [Figs.~\ref{fig:Gamma_dep}(b,d)]. This is because the relevant energy scale of the spin pumping is the exchange interaction, which is of the order of $\sim 1\ \mathrm{eV}$. In contrast, the SOC is necessary for the orbital pumping, whose energy scale is an order of magnitude smaller than the exchange interaction. 

The experimental work by Hayashi \emph{et al.} measured the DC component of the orbital pumping~\cite{Hayashi2023b}, which corresponds to $\chi_{xx}^L w \sin^2 \theta_\mathrm{c}$ in our theory. They found that the orbital pumping is significant in Ti/Ni while the effect is nearly absent in Ti/Fe, which seems consistent with our theoretical calculation. However, the numerical value of $\chi_{xx}^L$ for Ni is approximately three times larger than that for Fe, which cannot explain almost complete suppression of the signal in Ti/Fe samples. Therefore, we conclude that the orbital pumping in the bulk ferromagnet alone cannot explain the experimental result. We speculate that the OAM transmission or transparency at the Ti/Fe interface could be much lower than that at the Ti/Ni interface. 

Recent terahertz spectroscopy experiments have revealed that ferromagnetic Ni can generate the OAM by laser extiation~\cite{Wang2023, Seifert2023, Xu2023a}. Although the dynamics of magnetic moments is incoherent, which is different from the ferromagnetic resonance, we remark that the DC component projected onto the precession axis does not cancel as it is phase-independent~(Fig.~\ref{fig:orbital_pumping_schematics}). Thus, our results imply that  excitation of incoherent magnons by laser or temperature can induce OAM. In particular, we predict that when the gradient of such excitation is present, so-called \emph{orbital Seebeck effect} may appear by a similar mechanism of the orbital pumping. Recently, El Hamdi \emph{et al.} have shown that a temperature gradient can generate orbital currents, which is electrically measured via the inverse orbital Edelstein effect~\cite{ElHamdi2023}. The experiment suggests that 6 percents of the OAM is injected compared to the spin for ferromagnetic Permalloy, which is in reasonable agreement with our calculation (Tab.~\ref{tab:ratio}). In ferromagnets with low crystal symmetry, another mechanism that enables to induce the OAM from the thermal fluctuation of magnetic moments is via magnon-mediated chirality~\cite{Zhang2020}, which can result in an anomalous behavior of the $g$-factor as a function of the temperature~\cite{Alahmed2022}. The OAM can also be generated by combining the spin pumping and the spin-to-orbital conversion, which can be achieved by adding an insertion layer with strong SOC such as Pt~\cite{Santos2023, Xu2023b}.

In conclusion, we have developed a theoretical formalism that describes the orbital pumping in ferromagnets. Unlike the spin pumping, the orbital pumping crucially depends on the SOC. We have performed first-principles calculations on ferromagnetic Fe, Co, and Ni for quantitative estimation of the orbital pumping. The results shows that the orbital pumping is strongest in Ni and weakest in Fe, whose magnitude ranges from 5 to 15 percents compared to the magnitude of the spin pumping. Our work opens the possibility of generating the OAM by magnetization dynamics and serves as a guideline for experiments. 

{\it Note added} -- During the preparation of the submission of our manuscript, which was posted on arXiv in September 2023~\cite{Go2023}, we have noticed a similar work as ours on the theory of orbital pumping~\cite{Han2023}. We note that Ref.~\cite{Han2023} does not include quantitative estimation of the orbital pumping in real materials nor discussion on the role of microscopic electronic structure, which are part of the main results of our work.

D.G. acknowledges stimulating discussion with Michel Viret, Cosimo Gorini, and Giovanni Vignale. This work was funded by the Deutsche Forschungsgemeinschaft (DFG, German Research Foundation) $-$ TRR 173 $-$ 268565370 (project A11), TRR 288 $-$ 422213477 (project B06). K.A. acknowledges the support by JSPS KAKENHI (Grant Number: 22H04964, 20H00337, 20H02593), Spintronics Research Network of Japan (Spin-RNJ), and MEXT Initiative to Establish Next-generation Novel Integrated Circuits Centers (X-NICS) (Grant Number: JPJ011438). A.P. acknowledges support from the ANR ORION project, grant ANR-20-CE30-0022-01 of the French Agence Nationale de la Recherche. A. M. acknowledges support from the Excellence Initiative of Aix-Marseille Universit\'e - A*Midex, a French ``Investissements d’Avenir'' program. We  also gratefully acknowledge the J\"ulich Supercomputing Centre and RWTH Aachen University for providing computational resources under projects  cjiff40 and jara0062.

\let\oldaddcontentsline\addcontentsline
\renewcommand{\addcontentsline}[3]{}

\let\addcontentsline\oldaddcontentsline

\clearpage 
\widetext

\setcounter{equation}{0}
\setcounter{figure}{0}
\setcounter{table}{0}
\setcounter{page}{1}

\renewcommand{\theequation}{S\arabic{equation}}
\renewcommand{\thefigure}{S\arabic{figure}}
\renewcommand{\bibnumfmt}[1]{[S#1]}
\renewcommand{\citenumfont}[1]{S#1}
\renewcommand{\thepage}{S\arabic{page}}

\makeatletter
\def\@hangfrom@section#1#2#3{\@hangfrom{#1#2}#3}
\def\@hangfroms@section#1#2{#1#2}
\makeatother

\makeatletter
\renewcommand*{\thesection}{\arabic{section}}
\renewcommand*{\thesubsection}{\thesection.\arabic{subsection}}
\renewcommand*{\p@subsection}{}
\renewcommand*{\thesubsubsection}{\thesubsection.\arabic{subsubsection}}
\renewcommand*{\p@subsubsection}{}
\makeatother

\linespread{1.0}

\begin{center}
\Large \bf    
Supplemental Material for
\\
``Orbital Pumping by Magnetization Dynamics in Ferromagnets''
\end{center}

\begin{center}
{Dongwook Go, Kazuya Ando, Armando Pezo, Stefan Bl\"ugel, Aur\'elien Manchon, and Yuriy Mokrousov}
\end{center}

\tableofcontents

\section{Derivation of the orbital pumping response}

\subsection{Assumptions}

We start from the adiabatic perturbation theory~\cite{Vanderbilt2018} by assuming that the dynamics of the magnetization is sufficiently slower compared to the inverse of the characteristic energy difference between two arbitrary electronic states. In this case, an electronic eigenstate $\ket{n}$ evolves adiabatically by following the time-dependent change of the Hamiltonian without making a transition to another state, satisfying the eigenvalue equation
\begin{eqnarray}
\label{eq:eigenvalue_equation}
\mathcal{H}(t) \ket{n(t)} = \mathcal{E}_n (t) \ket{n(t)}.
\end{eqnarray}
Here, $\mathcal{H}(t)$ is the Hamiltonian, $\ket{n(t)}$ is the eigenstate, and $\mathcal{E}_n (t)$ is the energy eigenvalue at time $t$. The time-derivative of the expectation value of the orbital angular momentum (OAM) $\mathbf{L}$ is generally given by
\begin{equation}
\label{eq:partial_L_first}
\partial_t 
\left\langle 
\mathbf{L}
\right\rangle
=
\sum_n 
\partial_t f (\mathcal{E}_n)
\bra{n} \mathbf{L} \ket{ n}
+
\sum_n 
f(\mathcal{E}_n)
\left[ 
\bra{\partial_t  n} \mathbf{L} \ket{n}
+
\bra{n} \mathbf{L} \ket{\partial_t n}
\right],
\end{equation}
where $f (\mathcal{E}_n)$ is the Fermi-Dirac distribution function for the energy $\mathcal{E}_n$. The first term on the right hand side of Eq.~\eqref{eq:partial_L_first} captures the Fermi surface breathing effect, which is unique in metals and absent in insulators. Let us denote this term \emph{occupation-change} contribution. The second term is due to the adiabatic change of the eigenstates, which we denote by \emph{state-change} contribution.

For the microscopic Hamiltonian, we have the following assumptions. First, only the spin $\mathbf{S}$ couples to the magnetization in the following form of the Hamiltonian,
\begin{eqnarray}
\label{eq:Hamiltonian}
\mathcal{H}(t) =  \mathcal{H}_0 + J(\mathbf{r})\hat{\mathbf{m}}(t)\cdot\mathbf{S},
\end{eqnarray}
where the second term is the exchange interaction between the magnetization and the spin, and the rest of the Hamiltonian is indicated by $\mathcal{H}_0$. Second, we assume that the exchange coupling $J(\mathbf{r})$ does not depend on $t$ nor the direction of the magnetization. That is, magnetization dynamics occurs such that only the direction varies while the exchange field strength remains constant. This may be justified if the cone angle of the magnetization precession in the ferromagnetic resonance is small enough. Moreover, for elementary $3d$ ferromganets, the magnitude of the magnetization exhibits little angular dependence. Finally, we assume that the magnetization is uniform in space, i.e. no domain or real-space textures extending beyond the chemical unit cell. However, we allow the possibility that the exchange coupling $J(\mathbf{r})$ is position-dependent. This is the case for the spin-dependent density functional theory (DFT), which we use for the first-principles calculation. We emphasize that in principle, our formalism can be applied to real-space magnetic textures if the entire magnetic texture is included in the unit cell.

With the Hamiltonian in Eq.~\eqref{eq:Hamiltonian}, we define \emph{exchange torque} as
\begin{eqnarray}
T_\mathrm{XC}^{\mathbf{S}}
=
\frac{1}{i\hbar}
\left[ 
\mathbf{S}, \ J \hat{\mathbf{m}}\cdot \mathbf{S}
\right]
=
J \hat{\mathbf{m}}\times \mathbf{S},
\end{eqnarray}
which describes the transfer of angular momentum between the magnetization (order parameter) and the spin of electrons. This term keep appearing in describing the time-derivative of the Hamiltonian
\begin{eqnarray}
\label{eq:partial_H}
\partial_t \mathcal{H}
=
J \partial_t\hat{\mathbf{m}} \cdot\mathbf{S}
=
\left ( J \hat{\mathbf{m}}\times \mathbf{S} \right ) 
\cdot
\left ( \hat{\mathbf{m}}\times \partial_t \hat{\mathbf{m}} \right )
=
T_\mathrm{XC}^\mathbf{S}\cdot 
\left ( \hat{\mathbf{m}}\times \partial_t \hat{\mathbf{m}} \right )
,
\end{eqnarray}
which is repeatedly used in the derivation below. Note the vector $\hat{\mathbf{m}}\times \partial_t \hat{\mathbf{m}}$ pointing in the direction of the Gilbert damping naturally appears. 

\subsection{Occupation-change contribution}

The occupation-change contribution to the orbital pumping, the first term on the right hand side of Eq.~\eqref{eq:partial_L_first}, can be expressed as
\begin{eqnarray}
\partial_t \left\langle \mathbf{L} \right\rangle_\mathrm{occu}
&=&
\sum_n 
f'(\mathcal{E}_n) 
\bra{n} \mathbf{L} \ket{n} (\partial_t \mathcal{E}_n) 
\nonumber 
\\
&=&
\sum_n 
f'(\mathcal{E}_n) 
\bra{n} \mathbf{L} \ket{n}
\bra{n} \partial_t \mathcal{H} \ket{n}
\nonumber
\\
&=&
\sum_n 
f'(\mathcal{E}_n) 
\bra{n} \mathbf{L} \ket{n}
\bra{n} T_\mathrm{XC}^\mathbf{S} \ket{n}
\cdot 
\left ( \hat{\mathbf{m}}\times \partial_t \hat{\mathbf{m}} \right ),
\label{eq:occupation_change_first}
\end{eqnarray}
where $f'(\mathcal{E}_n) = \partial_{\mathcal{E}_n} f (\mathcal{E}_n)$ is the energy derivative of the Fermi-Dirac distribution function. On the second line, we have made use of the force theorem, and on the third line, we have employed Eq.~\eqref{eq:partial_H}. If we rewrite Eq.~\eqref{eq:occupation_change_first} as a response form, it becomes
\begin{eqnarray}
\partial_t \left\langle L_\alpha \right\rangle_\mathrm{occu}
=
\sum_\beta 
\chi_{\alpha\beta,\mathrm{occu}}^{L}
\left ( \hat{\mathbf{m}}\times \partial_t \hat{\mathbf{m}} \right )_\beta,
\end{eqnarray}
where $\alpha$ and $\beta$ are Cartesian indices, and 
\begin{eqnarray}
\chi_{\alpha\beta,\mathrm{occu}}^{L}
=
\sum_n 
f'(\mathcal{E}_n) 
\bra{n} L_\alpha \ket{n}
\bra{n} T_\mathrm{XC}^{S_\beta} \ket{n}.
\label{eq:chi_occupation}
\end{eqnarray}

\subsection{State-change contribution}

The state-change contribution, the second term on the right hand side of Eq.~\eqref{eq:partial_L_first}, can be written as
\begin{eqnarray}
\partial_t 
\left\langle 
\mathbf{L}
\right\rangle_\mathrm{state}
&=&
\sum_{n}
f(\mathcal{E}_n)
\sum_{m\neq n}
\left[ 
\braket{\partial_t n | m} 
\bra{m}\mathbf{L} \ket{n} 
+
\bra{n} \mathbf{L} \ket{m} 
\braket{m|\partial_t n}
\right]
\nonumber 
\\
&=&
\sum_{n}
\sum_{m\neq n}
\left[ 
f(\mathcal{E}_n) - f(\mathcal{E}_m)
\right]
\left[ 
\bra{n} \mathbf{L} \ket{m} 
\braket{m|\partial_t n}
\right]
\label{eq:state_change_2}
\end{eqnarray}
On the first line, $n=m$ contribution vanishes because $
\braket{n|\partial_t n}=-\braket{\partial_t n| n}$. On the second line, we have used $
\braket{m|\partial_t n}=-\braket{\partial_t m| n}$ and
exchanged the dummy variables $n$ and $m$, where it can also be seen that $n=m$ contribution vanishes due to the factor $f(\mathcal{E}_n)-f(\mathcal{E}_m)$, thus consistent with the first line.

We use the identity,
\begin{eqnarray}
\ket{\partial_t n}
=
\ket{n}
\braket{n|\partial_t n}
+
\sum_{m\neq n}
\frac{
\ket{m} \bra{m} \partial_t \mathcal{H} \ket{n}
}{
\mathcal{E}_n - \mathcal{E}_m + i\eta 
},
\end{eqnarray}
which can be easily shown by taking the time-derivative of Eq.~\eqref{eq:eigenvalue_equation}. Here we have inserted an infinitesimally small variable $\eta>0$ to ensure that the contribution from degenerate energies ($\mathcal{E}_n=\mathcal{E}_m$ but $n\neq m$) does not blow up to infinity, which will be taken to the limit $\eta \rightarrow 0+$ at the end of the calculation. Thus,  Eq.~\eqref{eq:state_change_2} becomes
\begin{eqnarray}
\partial_t 
\left\langle 
\mathbf{L}
\right\rangle_\mathrm{state}
&=&
\sum_{n}
\sum_{m\neq n}
\left[ 
f(\mathcal{E}_n) - f(\mathcal{E}_m)
\right]
\mathrm{Re}
\left[
\frac{
\bra{n} \mathbf{L} \ket{m} 
\bra{m} \partial_t \mathcal{H} \ket{n}
}{
\mathcal{E}_n - \mathcal{E}_m + i\eta 
}
\right].
\label{eq:state_change_3}
\end{eqnarray}
Finally, by using Eq.~\eqref{eq:partial_H}, the state-change contribution can be written as a perturbation-like expression,
\begin{eqnarray}
\partial_t 
\left\langle 
\mathbf{L}
\right\rangle_\mathrm{state}
&=&
\sum_{n}
\sum_{m\neq n}
\left[ 
f(\mathcal{E}_n) - f(\mathcal{E}_m)
\right]
\mathrm{Re}
\left[ 
\frac{
\bra{n} \mathbf{L} \ket{m} 
\bra{m}  T_\mathrm{XC}^\mathbf{S} \ket{n}
}{
\mathcal{E}_n - \mathcal{E}_m + i\eta 
}
\right]
\cdot
\left ( \hat{\mathbf{m}}\times \partial_t \hat{\mathbf{m}} \right ),
\label{eq:state_change_4}
\end{eqnarray}
Note the similarity between Eqs.~\eqref{eq:occupation_change_first} and \eqref{eq:state_change_4}.

Because the contribution by degenerate pairs ($\mathcal{E}_n=\mathcal{E}_m$) is excluded by the factor $f(\mathcal{E}_n) - f(\mathcal{E}_m)$, we can use the identity
\begin{eqnarray}
\label{eq:identity_1}
\bra{n} \mathbf{L} \ket{m}
=
-\frac{i\hbar}{\mathcal{E}_n - \mathcal{E}_m + i\eta }
\bra{n} T^\mathbf{L}_\mathrm{tot} \ket{m},
\end{eqnarray}
where 
\begin{eqnarray}
T^\mathbf{L}_\mathrm{tot} = \frac{1}{i\hbar}
\left[ 
\mathbf{L}, \ \mathcal{H}
\right]
\end{eqnarray}
is the total orbital torque. Thus, Eq.~\eqref{eq:state_change_4} becomes
\begin{eqnarray}
\partial_t 
\left\langle 
\mathbf{L}
\right\rangle_\mathrm{state}
&=&
\hbar 
\sum_{n}
\sum_{m\neq n}
\left[ 
f(\mathcal{E}_n) - f(\mathcal{E}_m)
\right]
\mathrm{Im}
\left[ 
\frac{
\bra{n} T_\mathrm{tot}^\mathbf{L} \ket{m} 
\bra{m}  T_\mathrm{XC}^\mathbf{S} \ket{n}
}{
(\mathcal{E}_n - \mathcal{E}_m + i\eta )^2
}
\right]
\cdot
\left ( \hat{\mathbf{m}}\times \partial_t \hat{\mathbf{m}} \right ).
\label{eq:state_change_5}
\end{eqnarray}

Finally, we use a cute trick from the observation that the contribution $n=m$ does not contribute anyway in Eq.~\eqref{eq:state_change_5}, which can be included in the sum. Therefore, we have
\begin{eqnarray}
\partial_t 
\left\langle 
L_\alpha 
\right\rangle_\mathrm{state}
&=&
\sum_\beta 
\chi_{\alpha\beta,\mathrm{state}}^L
\left ( \hat{\mathbf{m}}\times \partial_t \hat{\mathbf{m}} \right )_\beta,
\end{eqnarray}
where
\begin{eqnarray}
\chi_{\alpha\beta,\mathrm{state}}^L
=
\hbar 
\sum_{n}
\sum_{m}
\left[ 
f(\mathcal{E}_n) - f(\mathcal{E}_m)
\right]
\mathrm{Im}
\left[ 
\frac{
\bra{n} T_\mathrm{tot}^{L_\alpha} \ket{m} 
\bra{m}  T_\mathrm{XC}^{S_\beta} \ket{n}
}{
(\mathcal{E}_n - \mathcal{E}_m + i\eta )^2
}
\right].
\label{eq:state_change_final}
\end{eqnarray}

\subsection{Green's function representation}

Here, we show that the state-change contribution [Eq.~\eqref{eq:state_change_final}] can be represented in terms of Green's functions. We adopt several tricks for manipulating equations, as presented in Ref.~\cite{Crepieux2001} in the derivation of the Bastin-Smr\v{c}ka-St\v{r}eda's form~\cite{Bastin1971,Smrcka1977} of the Kubo formula~\cite{Kubo1956}. The starting point is to insert a ``magic number 1'', which equals the integral of the Dirac delta function, into Eq.~\eqref{eq:state_change_final},
\begin{eqnarray}
\chi_{\alpha\beta,\mathrm{state}}^L
& & =
\hbar 
\int d\mathcal{E}
\sum_{n}
f(\mathcal{E})
\delta (\mathcal{E}-\mathcal{E}_n)
\mathrm{Im}
\left[ 
\bra{n} T_\mathrm{tot}^{L_\alpha}
\sum_{m}
\left( 
\frac{
\ket{m} 
\bra{m}
}{
\mathcal{E} - \mathcal{E}_m + i\eta
}
\right)^2
T_\mathrm{XC}^{S_\beta} \ket{n}
\right]
\nonumber 
\\
& &
-
\hbar 
\int d\mathcal{E}
\sum_{m}
f(\mathcal{E})
\delta (\mathcal{E}-\mathcal{E}_m)
\mathrm{Im}
\left[ 
\bra{m}  T_\mathrm{XC}^{S_\beta}
\sum_{n}
\left( 
\frac{
 \ket{n}
\bra{n} 
}{
\mathcal{E} - \mathcal{E}_n - i\eta 
}
\right)^2
T_\mathrm{tot}^{L_\alpha} \ket{m} 
\right].
\label{eq:Green_function_representation_step_1}
\end{eqnarray}
In the Lehmann representation for the exact eigenstates, the retarded/advanced Green's function is given by
\begin{equation}
G^{\mathrm{R/A}} (\mathcal{E})
=
\sum_n
\ket{n} G_n^\mathrm{R/A} (\mathcal{E}) \bra{n},
\end{equation}
where 
\begin{equation}
 G_n^\mathrm{R/A}(\mathcal{E}) = \frac{1}{\mathcal{E} - \mathcal{E}_n \pm i\eta},
\end{equation}
and 
\begin{equation}
\partial_\mathcal{E} G_n^\mathrm{R/A}(\mathcal{E}) = - \left[ 
G_n^\mathrm{R/A}(\mathcal{E})
\right]^2.
\end{equation}
Also, the Dirac delta function, which is proportional to the spectral function, can be written in terms of the exact Green's functions,
\begin{equation}
\delta (\mathcal{E} - \mathcal{E}_n)
=
\frac{i}{2\pi}
\left[ 
G_n^\mathrm{R} (\mathcal{E}) - G_n^\mathrm{A} (\mathcal{E})
\right].
\end{equation}
By using these identities, Eq.~\eqref{eq:Green_function_representation_step_1} can be written as
\begin{eqnarray}
\chi_{\alpha\beta,\mathrm{state}}^L
&=&
- \frac{\hbar}{2\pi} 
\int d\mathcal{E}
\sum_{n}
f(\mathcal{E})
\mathrm{Re}
\left[ 
\left\{
G_n^\mathrm{R} (\mathcal{E}) - G_n^\mathrm{A} (\mathcal{E})
\right\}
\bra{n} T_\mathrm{tot}^{L_\alpha}
\sum_{m}
\ket{m}
\partial_\mathcal{E}
G^\mathrm{R}_m (\mathcal{E})
\bra{m}
T_\mathrm{XC}^{S_\beta} \ket{n}
\right]
\nonumber 
\\
& &
+
\frac{\hbar}{2\pi} 
\int d\mathcal{E}
\sum_{m}
f(\mathcal{E})
\mathrm{Re}
\left[ 
\left\{
G_m^\mathrm{R} (\mathcal{E}) - G_m^\mathrm{A} (\mathcal{E})
\right\}
\bra{m}  T_\mathrm{XC}^{S_\beta}
\sum_{n}
\ket{n}
\partial_\mathcal{E}
G^\mathrm{A}_n (\mathcal{E})
\bra{n}
T_\mathrm{tot}^{L_\alpha} \ket{m} 
\right].
\nonumber 
\\
\end{eqnarray}
Finally, this can be written in a gauge-independent manner without band indices,
\begin{eqnarray}
\chi_{\alpha\beta,\mathrm{state}}^L
=& & 
\frac{\hbar}{2\pi} 
\int d\mathcal{E}
f(\mathcal{E})
\mathrm{Tr}
\left[ 
T_\mathrm{tot}^{L_\alpha}
\left\{ 
G^R (\mathcal{E}) - G^A (\mathcal{E})
\right\}
T_\mathrm{XC}^{S_\beta}
\partial_\mathcal{E} G^A (\mathcal{E})
\right.
\nonumber 
\\
& & 
\ \ \ \ \ \ \ \ \ \ \ \ \ \ \ \ \ \ \ \ \ \ \ \ \ \ 
-
\left. 
T_\mathrm{tot}^{L_\alpha}
\partial_\mathcal{E} G^R (\mathcal{E})
T_\mathrm{XC}^{S_\beta}
\left\{ 
G^R (\mathcal{E}) - G^A (\mathcal{E})
\right\}
\right],
\label{eq:Green_function_representation_step_2}
\end{eqnarray}
where the cyclic invariance of the trace has been used. Equation \eqref{eq:Green_function_representation_step_2} is in fact identical to Eq.~(2) of the main text. The time-reversal-even contribution [Eq.~(2a) in the main text]
\begin{eqnarray}
\chi_{\alpha\beta,\mathrm{even}}^L
& & = 
\frac{\hbar}{4\pi}
\int d\mathcal{E} 
\left\{ 
\partial_\mathcal{E} f(\mathcal{E})
\right\}
\mathrm{Tr}
\left[
T_\mathrm{tot}^{L_\alpha}
\left\{ 
G^\mathrm{R}(\mathcal{E}) - G^\mathrm{A} (\mathcal{E})
\right\}
T_\mathrm{XC}^{S_\beta}
\left\{ 
G^\mathrm{R}(\mathcal{E}) - G^\mathrm{A} (\mathcal{E})
\right\}
\right],
\label{eq:chi_even}
\end{eqnarray}
can be re-written using integration-by-part,
\begin{eqnarray}
\chi_{\alpha\beta,\mathrm{even}}^L
& & =
-\frac{\hbar}{4\pi}
\int d\mathcal{E} 
f(\mathcal{E})
\mathrm{Tr}
\left[
T_\mathrm{tot}^{L_\alpha}
\left\{ 
\partial_\mathcal{E} G^\mathrm{R}(\mathcal{E}) - \partial_\mathcal{E} G^\mathrm{A} (\mathcal{E})
\right\}
T_\mathrm{XC}^{S_\beta}
\left\{ 
G^\mathrm{R}(\mathcal{E}) - G^\mathrm{A} (\mathcal{E})
\right\}
\right. 
\nonumber 
\\
& & 
\ \ \ \ \ \ \ \ \ \ \ \ \ \ \ \ \ \ \ \ \ \ \ \ \ \ \ \ \ \ \ \ \ 
+
\left. 
T_\mathrm{tot}^{L_\alpha}
\left\{ 
G^\mathrm{R}(\mathcal{E}) - G^\mathrm{A} (\mathcal{E})
\right\}
T_\mathrm{XC}^{S_\beta}
\left\{ 
\partial_\mathcal{E} G^\mathrm{R}(\mathcal{E}) - \partial_\mathcal{E} G^\mathrm{A} (\mathcal{E})
\right\}
\right].
\end{eqnarray}
If we sum this with the time-reversal-odd contribution [Eq.~(2b) in the main text]
\begin{eqnarray}
\chi_{\alpha\beta,\mathrm{odd}}^L
= & & 
\frac{\hbar}{4\pi}
\int d\mathcal{E} f (\mathcal{E})
\mathrm{Tr}
\left[ 
T_\mathrm{tot}^{L_\alpha}
\left\{ G^\mathrm{R} (\mathcal{E})  - G^\mathrm{A} (\mathcal{E}) \right\}
T_\mathrm{XC}^{S_\beta}
\left\{
\partial_\mathcal{E} G^\mathrm{R} (\mathcal{E})  + \partial_\mathcal{E} G^\mathrm{A} (\mathcal{E})
\right\}
\right.
\nonumber
\\
& &
\ \ \ \ \ \ \ \ \ \ \ \ \ \ \ \ \ \ \ \ \ \ \ \ \ \ 
\left.
-
T_\mathrm{tot}^{L_\alpha}
\left\{ 
\partial_\mathcal{E} G^\mathrm{R} (\mathcal{E}) + \partial_\mathcal{E} G^\mathrm{A} (\mathcal{E})
\right\}
T_\mathrm{XC}^{S_\beta}
\left\{ 
G^\mathrm{R} (\mathcal{E}) - G^\mathrm{A} (\mathcal{E})
\right\}
\right],
\label{eq:chi_odd}
\end{eqnarray}
we recover Eq.~\eqref{eq:Green_function_representation_step_2}. That is,
\begin{eqnarray}
\chi_{\alpha\beta,\mathrm{state}}^L = 
\chi_{\alpha\beta,\mathrm{even}}^L
+
\chi_{\alpha\beta,\mathrm{odd}}^L.
\end{eqnarray}

\subsection{Constant broadening approximation}

As an approximation, we assume the constant broadening $\Gamma > 0$ of the energy spectrum of single-electron Bloch states. Thus, the single-particle Green's function contains $\Gamma$ in the imaginary part of the self-energy,
\begin{equation}
 G_{n\mathbf{k}}^\mathrm{R/A}(\mathcal{E}) = \frac{1}{\mathcal{E} - \mathcal{E}_{n\mathbf{k}} \pm i\Gamma},
\end{equation}
where $n$ is the band index and $\mathbf{k}$ is the crystal momentum of the Bloch state $\psi_{n\mathbf{k}}$. We also assume the zero temperature in the Fermi-Dirac distribution function $f(\mathcal{E})=\Theta (\mathcal{E}_\mathrm{F}-\mathcal{E})$, where $\Theta (x)$ is the Heaviside step function and $\mathcal{E}_\mathrm{F}$ is the Fermi energy. Thus $\partial_\mathcal{E} f(\mathcal{E})=-\delta (\mathcal{E}-\mathcal{E}_\mathrm{F})$. With these assumptions, the energy integral can be analytically performed, and the explicit expressions for Eqs.~\eqref{eq:chi_even} and \eqref{eq:chi_odd} [Eqs.~(2a) and (2b) in the main text, respectively] become
\begin{eqnarray}
\chi_{\alpha\beta,\mathrm{even}}^L
&=&
\frac{\hbar}{\pi}
\sum_\mathbf{k}
\sum_{nm}
\mathrm{Re}
\left[ 
\bra{\psi_{n\mathbf{k}}}
T_\mathrm{tot}^{L_\alpha}
\ket{\psi_{m\mathbf{k}}}
\bra{\psi_{m\mathbf{k}}}
T_\mathrm{XC}^{S_\beta}
\ket{\psi_{n\mathbf{k}}}
\right]
A_{n\mathbf{k}} (\mathcal{E}_\mathrm{F})
A_{m\mathbf{k}} (\mathcal{E}_\mathrm{F})
,
\label{eq:chi_even_implemented}
\end{eqnarray}
and
\begin{eqnarray}
\chi_{\alpha\beta,\mathrm{odd}}^L
&=&
\frac{\hbar}{\pi}
\sum_\mathbf{k}
\sum_{nm}
\mathrm{Im}
\left[ 
\bra{\psi_{n\mathbf{k}}}
T_\mathrm{tot}^{L_\alpha}
\ket{\psi_{m\mathbf{k}}}
\bra{\psi_{m\mathbf{k}}}
T_\mathrm{XC}^{S_\beta}
\ket{\psi_{n\mathbf{k}}}
\right]
W_{nm\mathbf{k}}(\mathcal{E}_\mathrm{F})
,
\label{eq:chi_odd_implemented}
\end{eqnarray}
where 
\begin{eqnarray}
W_{nm\mathbf{k}}(\mathcal{E}_\mathrm{F})
&=&
-\frac{1}{(\mathcal{E}_{n\mathbf{k}} - \mathcal{E}_{m\mathbf{k}})^2}
\left[ 
\arctan \left( 
\frac{\mathcal{E}_{n\mathbf{k}}-\mathcal{E}_\mathrm{F}}{\Gamma}
\right)
-
\arctan \left( 
\frac{\mathcal{E}_{m\mathbf{k}}-\mathcal{E}_\mathrm{F}}{\Gamma}
\right)
\right]
\nonumber
\\
& & + 
\frac{\Gamma}{\mathcal{E}_{n\mathbf{k}} - \mathcal{E}_{m\mathbf{k}}}
\left[ 
A_{n\mathbf{k}}(\mathcal{E}_\mathrm{F}) \bar{A}_{m\mathbf{k}}(\mathcal{E}_\mathrm{F})
+\bar{A}_{n\mathbf{k}}(\mathcal{E}_\mathrm{F}) {A}_{m\mathbf{k}}(\mathcal{E}_\mathrm{F})
\right],
\end{eqnarray}
\begin{eqnarray}
A_{n\mathbf{k}} (\mathcal{E}_\mathrm{F})
=
\frac{i}{2}
\left[ 
G_{n\mathbf{k}}^\mathrm{R} (\mathcal{E}_\mathrm{F}) - G_{n\mathbf{k}}^\mathrm{A} (\mathcal{E}_\mathrm{F})
\right]
=
\frac{\Gamma}{
(\mathcal{E}_{n\mathbf{k}}-\mathcal{E}_\mathrm{F})^2 + \Gamma^2
},
\end{eqnarray}
and 
\begin{eqnarray}
\bar{A}_{n\mathbf{k}} (\mathcal{E}_\mathrm{F})
=
\frac{
\mathcal{E}_{n\mathbf{k}}-\mathcal{E}_\mathrm{F}
}{
(\mathcal{E}_{n\mathbf{k}}-\mathcal{E}_\mathrm{F})^2 + \Gamma^2
}.
\end{eqnarray}
We remark that Eqs.~\eqref{eq:chi_even_implemented} and \eqref{eq:chi_odd_implemented} are the formulae which we have numerically implemented for the first-principles calculation. Meanwhile, note that the occupation-change contribution [Eq.~\eqref{eq:occupation_change_first}] does not depend on the broadening, thus
\begin{eqnarray}
\chi_{\alpha\beta,\mathrm{occu}}^{L}
=
\sum_\mathbf{k}
\sum_n 
f'(\mathcal{E}_{n\mathbf{k}}) 
\bra{\psi_{n\mathbf{k}}} L_\alpha \ket{\psi_{n\mathbf{k}}}
\bra{\psi_{n\mathbf{k}}} T_\mathrm{XC}^{S_\beta} \ket{\psi_{n\mathbf{k}}}.
\label{eq:chi_occupation_final}
\end{eqnarray}

\section{Reciprocal relation between the orbital torque and pumping}

The orbital torque [Eq.~(3) in the main text] is the response of the exchange torque $T_\mathrm{XC}^\mathbf{S}$ by the orbital-dependent Zeeman interaction,
\begin{eqnarray}
\mathcal{H}'= \mathbf{B}^L\cdot\mathbf{L},
\label{eq:OT_perturbation}
\end{eqnarray}
where $\mathbf{B}^L$ is an effective magnetic field having the dimension of $\text{[time]}^{-1}$. In the first-order perturbation, Eq.~\eqref{eq:OT_perturbation} has two consequences. The first consequence is the change of the occupation function, and the second consequence is the change of the quantum states. Thus, the expectation value of the exchange torque changes by
\begin{eqnarray}
\delta 
\left\langle 
T_\mathrm{XC}^{S_\alpha} 
\right\rangle
=
\sum_n f' (\mathcal{E}_n) \delta \mathcal{E}_n
\bra{n} T_\mathrm{XC}^{S_\alpha} \ket{n}
+
\sum_n 
f (\mathcal{E}_n)
\left[ 
\bra{\delta n}  T_\mathrm{XC}^{S_\alpha} \ket{n}
+
\bra{n}  T_\mathrm{XC}^{S_\alpha} \ket{\delta n}
\right],
\label{eq:OT_starting_point}
\end{eqnarray}
where $\delta \mathcal{E}_n$ and $\ket{\delta n}$ are the changes of the energy eigenvalue and the state ket. Note the similarity with Eq.~\eqref{eq:partial_L_first}. By the first-order perturbation theory, we have
\begin{eqnarray}
\delta \mathcal{E}_n
=
\bra{n} \mathcal{H}' \ket{n}
=
\sum_\beta  
\bra{n} L_\beta  \ket{n} B_\beta^L,
\label{eq:OT_energy_perturbation}
\end{eqnarray}
and 
\begin{eqnarray}
\ket{\delta n}
=
\sum_m
\frac{
\ket{m}\bra{m} \mathcal{H}' \ket{n}
}{
\mathcal{E}_n - \mathcal{E}_m +i\eta 
}
=
\sum_\beta 
\sum_m
\frac{
\ket{m}\bra{m} L_\beta \ket{n}
}{
\mathcal{E}_n - \mathcal{E}_m + i\eta 
} B_\beta^L.
\label{eq:OT_state_perturbation}
\end{eqnarray}

Thus, we may define occupation-change and state-change contributions analogous to Eq.~\eqref{eq:partial_L_first} on the orbital pumping,
\begin{eqnarray}
\delta 
\left\langle 
T_\mathrm{XC}^{S_\alpha} 
\right\rangle
=
\sum_\beta 
\left( 
\widetilde{\chi}_{\alpha\beta,\mathrm{occu}}^L
+
\widetilde{\chi}_{\alpha\beta,\mathrm{state}}^L
\right)
B_\beta^L,
\end{eqnarray}
where note the tilde sign that distinguishes the response tensors for the orbital pumping. By inserting Eqs.~\eqref{eq:OT_energy_perturbation} and \eqref{eq:OT_state_perturbation} into Eq.~\eqref{eq:OT_starting_point}, we obtain
\begin{eqnarray}
\widetilde{\chi}_{\alpha\beta,\mathrm{occu}}^L
=
\sum_n 
\bra{n} T_\mathrm{XC}^{S_\alpha} \ket{n}
\bra{n} L_\beta \ket{n},
\label{eq:OT_occupation_change}
\end{eqnarray}
and 
\begin{subequations}
\begin{eqnarray}
\widetilde{\chi}_{\alpha\beta,\mathrm{state}}^L
&=&
\sum_n 
\sum_{m\neq n}
\left[ 
f(\mathcal{E}_n) - f(\mathcal{E}_m)
\right]
\mathrm{Re}
\left[ 
\frac{
\bra{n} T_\mathrm{XC}^{S_\alpha} \ket{m}
\bra{m} L_\beta \ket{n}
}{
\mathcal{E}_n - \mathcal{E}_m + i\eta
}
\right]
\label{eq:OT_state_change_1}
\\
&=&
-
\hbar
\sum_n 
\sum_{m}
\left[ 
f(\mathcal{E}_n) - f(\mathcal{E}_m)
\right]
\mathrm{Im}
\left[ 
\frac{
\bra{n} T_\mathrm{XC}^{S_\alpha} \ket{m}
\bra{m} T_\mathrm{tot}^{L_\beta} \ket{n}
}{
(\mathcal{E}_n - \mathcal{E}_m + i\eta)^2
}
\right].
\label{eq:OT_state_change_2}
\end{eqnarray}
\end{subequations}
Note that in Eq.~\eqref{eq:OT_state_change_2}, we have used Eq.~\eqref{eq:identity_1} by replacing $+i\eta$ into $-i\eta$, which does not affect the final result after taking the limit $\eta \rightarrow 0+$. We also included $m=n$ contribution in the summation, which is zero due to the factor $f(\mathcal{E}_n) - f(\mathcal{E}_m)$.

Therefore, by comparing Eqs.~\eqref{eq:OT_occupation_change} and \eqref{eq:OT_state_change_2} with Eqs.~\eqref{eq:chi_occupation} and \eqref{eq:state_change_final}, we confirm the reciprocal relation for each occupation-change and state-change contribution:
\begin{subequations}
\begin{eqnarray}
\chi_{\alpha\beta,\mathrm{occu}}^L &=& \widetilde{\chi}_{\beta\alpha,\mathrm{occu}}^L,
\\
\chi_{\alpha\beta,\mathrm{state}}^L &=& \widetilde{\chi}_{\beta\alpha,\mathrm{state}}^L,
\end{eqnarray}
\end{subequations}
where the state-change contribution becomes identical in the limit $\eta \rightarrow 0+$. Note that the Green's function expression for the orbital pumping is also analogously written as Eqs.~\eqref{eq:chi_even} and \eqref{eq:chi_odd}. Thus, each of the even and odd part of the state-change contribution satisfies the reciprocal relations.

\section{Symmetry constraints on the orbital pumping}

Here, we examine the constraints imposed by symmetries on $\chi_{\alpha\beta,\mathrm{even}}^L$ [Eq.~\eqref{eq:chi_even_implemented}], $\chi_{\alpha\beta,\mathrm{odd}}^L$ [Eq.~\eqref{eq:chi_odd_implemented}], and $\chi_{\alpha\beta,\mathrm{occu}}^L$ [Eq.~\eqref{eq:chi_occupation_final}]. As examples, we examine Fe in the body-centered cubic (bcc), Co in the hexagonal close-packed (hcp), and Ni in the face-centered cubic (fcc) structures, which we consider for the first-principles calculation in Sec.~\ref{sec:first-principiles}. We assume $\hat{\mathbf{m}}=-\hat{\mathbf{z}}$ for the magnetization direction.

First, let us examine the behavior of $\chi_{\alpha\beta,\mathrm{even}}^L$, $\chi_{\alpha\beta,\mathrm{odd}}^L$, and $\chi_{\alpha\beta,\mathrm{occu}}^L$ under the time-reversal $\mathcal{T}$. Physically, this corresponds to the relation between the orbital pumpings in the original system and the time-reversed system in which the direction of the magnetization is opposite to that of the original system. This can reveal whether a particular contribution is an even or odd function of the magnetization. For evaluating the response function of the time-reversed system, we insert $\ket{\mathcal{T}n}$ in the place of $\ket{n}$. Here, one has to be careful for the anti-unitarity of $\mathcal{T}$. In general, for an arbitrary operator $A$ whose time-reversal parity is $\mathcal{P}_A$, the relation 
\begin{equation}
\bra{\mathcal{T}n} A \ket{\mathcal{T}m} = \mathcal{P}_A \bra{m} A \ket{n}
\label{eq:anti_unitary}
\end{equation}
holds. Because of an additional sign change on the imaginary part of the matrix element, one needs to pay a particular attention, in particular, when examining $\chi_{\alpha\beta,\mathrm{odd}}^L$ [Eq.~\eqref{eq:chi_odd_implemented}]. The time-reversal parities of angular momentum and torque operators are odd and even, respectively. Note that the time-reversal parity of the energy is even, which is in fact why the time-reversal is defined as an anti-unitary operator. Therefore, by Eq.~\eqref{eq:anti_unitary}, $\chi_{\alpha\beta,\mathrm{even}}^L$, one can show that $\chi_{\alpha\beta,\mathrm{odd}}^L$, and $\chi_{\alpha\beta,\mathrm{occu}}^L$ are even, odd, and odd functions of $\hat{\mathbf{m}}$.

Now let us consider spatial symmetry operations in combination with $\mathcal{T}$. When $\hat{\mathbf{m}}=-\hat{\mathbf{z}}$ in bcc, hcp, or fcc crystals, the following symmetries remain the system invariant,
\begin{eqnarray}
\label{eq:symmetry_list}
\mathcal{I}, \ \mathcal{M}_z,\ \mathcal{M}_x \mathcal{T},\ \mathcal{M}_y \mathcal{T},
\end{eqnarray}
where $\mathcal{I}$ is the spatial inversion, and $\mathcal{M}_\alpha$ is the mirror reflection with respect to $\alpha=x,y,z$ axis. Note that $\mathcal{M}_x \mathcal{T}$ and $\mathcal{M}_y \mathcal{T}$ are anti-unitary, which additionally changes the sign of the imaginary part of a matrix element as in Eq.~\eqref{eq:anti_unitary}. Before analyzing the constraints imposed by the symmetries in Eq.~\eqref{eq:symmetry_list}, let us make a table summarizing the parities;

\vspace{12pt}
\newcolumntype{P}[1]{>{\centering\arraybackslash}p{#1}}
\newcolumntype{M}[1]{>{\centering\arraybackslash}m{#1}}
\renewcommand{\arraystretch}{1.3}
\begin{table}[h!]
\begin{center}
\centering
\begin{tabular}{M{1.5cm} | M{1.0cm} M{1.0cm} M{1.0cm} | M{1.0cm} M{1.0cm} M{1.0cm} | M{2.0cm}}
\hline \hline
  & $L_x$ & $L_y$ & $L_z$ & $T_x$ & $T_y$ & $T_z$ & anti-unitary \\
\hline \hline 
$\mathcal{I}$ & $+$ & $+$ & $+$ & $+$ & $+$ & $+$ &  No \\ 
$\mathcal{M}_z$ & $-$ & $-$ & $+$ & $-$ & $-$ & $+$ &  No \\ 
$\mathcal{M}_x \mathcal{T}$ & $-$ & $+$ & $+$ & $+$ & $-$ & $-$ &  Yes  \\ 
$\mathcal{M}_y \mathcal{T}$ & $+$ & $-$ & $+$ & $-$ & $+$ & $-$ &  Yes  \\ 
\hline \hline
\end{tabular}
\end{center}
\end{table}

\noindent 
Note that unitary operators impose the same constraint for all $\chi_{\alpha\beta,\mathrm{even}}$, $\chi_{\alpha\beta,\mathrm{odd}}$, and $\chi_{\alpha\beta,\mathrm{occu}}$, which is not the case for anti-unitary operators $\mathcal{M}_x \mathcal{T}$ and $\mathcal{M}_y \mathcal{T}$. From this, one notices that  $\chi_{\alpha\beta,\mathrm{odd}}$ [Eq.~\eqref{eq:chi_odd_implemented}] and $\chi_{\alpha\beta,\mathrm{occu}}$ [Eq.~\eqref{eq:chi_occupation_final}] are subject to the same constraints.

Thus, each symmetry gives the following constraints. $\mathcal{I}$ does not impose any constraint. $\mathcal{M}_z$ forbids the components of $\chi_{\alpha\beta}$, in which only one of the indices contains $z$. That is, $\chi_{xz}$, $\chi_{zx}$, $\chi_{yz}$, and $\chi_{zy}$ are zero for all $\mathcal{T}$-even, $\mathcal{T}$-odd, and occupation-change contributions. From $\mathcal{M}_x \mathcal{T}$, the following components are zero: $\chi_{xy}$, $\chi_{xz}$, $\chi_{yx}$, $\chi_{zx}$ for the $\mathcal{T}$-even contribution; $\chi_{xx}$, $\chi_{yy}$, $\chi_{yz}$, $\chi_{zy}$, $\chi_{zz}$ for the $\mathcal{T}$-odd and occupation-change contributions. Similarly, $\mathcal{M}_y \mathcal{T}$ forbids following components: $\chi_{xy}$, $\chi_{yx}$, $\chi_{yz}$, $\chi_{zy}$ for the $\mathcal{T}$-even contribution; $\chi_{xx}$, $\chi_{xz}$, $\chi_{yy}$, $\chi_{zx}$, $\chi_{zz}$ for the $\mathcal{T}$-odd and occupation-change contributions. Therefore, combining all the symmetries leaves only the following nonzero components: $\chi_{xx}$, $\chi_{yy}$, $\chi_{zz}$ for the $\mathcal{T}$-even contribution; $\chi_{xy}$, $\chi_{yx}$ for the $\mathcal{T}$-odd and occupation-change contributions.

We remark that although $\chi_{zz}$ may be allowed, it is irrelevant because the generalized force $\hat{\mathbf{m}}\times \partial \hat{\mathbf{m}}$ is orthogonal to $\hat{\mathbf{z}}=-\hat{\mathbf{m}}$. Finally, the four-fold rotation symmetry in bcc/fcc crystals or six-fold rotation symmetry in hcp crystals with respect to $z$ axis imposes $\chi_{xx}=\chi_{yy}$ and $\chi_{xy}=-\chi_{yx}$. Therefore, in the first-principles calculation, we compute only $\chi_{xx}$ from the $\mathcal{T}$-even contribution and $\chi_{xy}$ from the $\mathcal{T}$-odd and occupation-change contributions.

\vspace{12pt}
\section{First-principles calculation of the orbital pumping}
\label{sec:first-principiles}

\subsection{Methods}

The first-principles calculation of the spin and orbital pumpings have been performed in the scheme of Wannier interpolation, consisting of three steps of calculations. As the first step, the DFT-based self-consistent calculation of the electronic structure has been carried out using  the \texttt{FLEUR} code~\cite{fleur}, in which the full-potential linearly augmented plane wave (FLAPW) method is implemented~\cite{Wimmer1981}. We emphasize that the FLAPW method allows direct access to the OAM; We explicitly integrate the matrix elements of operator $\mathbf{L}_\mathrm{atom}=(\mathbf{r}-\mathbf{r}_\mathrm{atom})\times\mathbf{p}$ in terms of the FLAPW basis function in the muffin-tin sphere, where $\mathbf{r}$ is the position operator, $\mathbf{r}_\mathrm{atom}$ is the position of an atom, and $\mathbf{p}$ is the momentum operator. For the exchange and correlation effects have been treated by using the Perdew-Burke-Ernzerhof functional that implements the generalized gradient approximation ~\cite{Perdew1996}. We have chosen the bcc ($a=5.42a_0$), hcp ($a=4,74a_0, \ c=7.68a_0$) and fcc ($a=6.65a_0$) structures for Fe, Co, and Ni, where the numbers in the parentheses indicate the lattice constants and $a_0$ is the Bohr radius. The muffin tin radii of Fe, Co, and Ni are all set $R_\mathrm{MT}=2.30a_0$, wherein the harmonics are expanded up to $l_\mathrm{max}=12$. In the interstitial region, the plane wave cutoff is set $K_\mathrm{max}=4.5a_0^{-1}$. We have additionally included local orbitals describing $3p$ states for better convergence of the spin density and potential. The spin-orbit coupling (SOC) is included in the second-variation scheme, where the spin quantization axis is set along the $c$-axis ($\hat{\mathbf{m}}=-\hat{\mathbf{z}}$).

In the second step, maximally localized Wannier functions (MLWFs) are obtained from the converged Kohn-Sham states in the first step, for which the \texttt{FLEUR} code is interfaced with the \texttt{WANNIER90} code~\cite{Freimuth2008,Pizzi2020}. We use $36 N_\mathrm{atom}$ Kohn-Sham states to construct $18 N_\mathrm{atom}$ MLWFs, where $N_\mathrm{atom}$ is the number of atoms in the unit cell ($N_\mathrm{atom}=1$ for Fe and Ni, and $N_\mathrm{atom}=2$ for Co). The 18 MLWFs are obtained from the initial projections of $s$, $p_x$, $p_y$, $p_z$, $d_{z^2}$, $d_{x^2-y^2}$, $d_{xy}$, $d_{yz}$, $d_{zx}$ shape Wannier states for spin up and down states. To ensure that $sd$ orbital characters are contained in the MWLFs, we set the frozen energy window in the disentanglement step, whose maximum is set $5\ \mathrm{eV}$ above the Fermi energy. Then we iteratively minimize the spread of the Wannier functions until they are converged. From the converged MWLFs, the Hamiltonian, spin and OAM operators, which are obtained first in the FLAPW basis in the first step, are transformed into the representation in terms of the MWLFs.

In the last step, from the obtained operators written in terms of the MLWFs in real space, we diagonalize the Hamiltonian and obtain the Bloch states and the energy eigenvalues and evaluate the formulae in Eqs.~\eqref{eq:chi_even_implemented} and \eqref{eq:chi_odd_implemented}. In this step, we use the in-house code \texttt{ORBITRANS}, which can compute orbital response properties in the Wannier representation in a highly efficient manner with massive parallelization in a hybrid scheme (the code is unavailable to the public yet but is to be released in the near future). 

\newpage 

\subsection{Negligible magnitude of the occupation-change contribution}

\begin{figure}[h!]
\vspace{12pt}
\centering
\includegraphics[angle=0, width=0.97\textwidth]{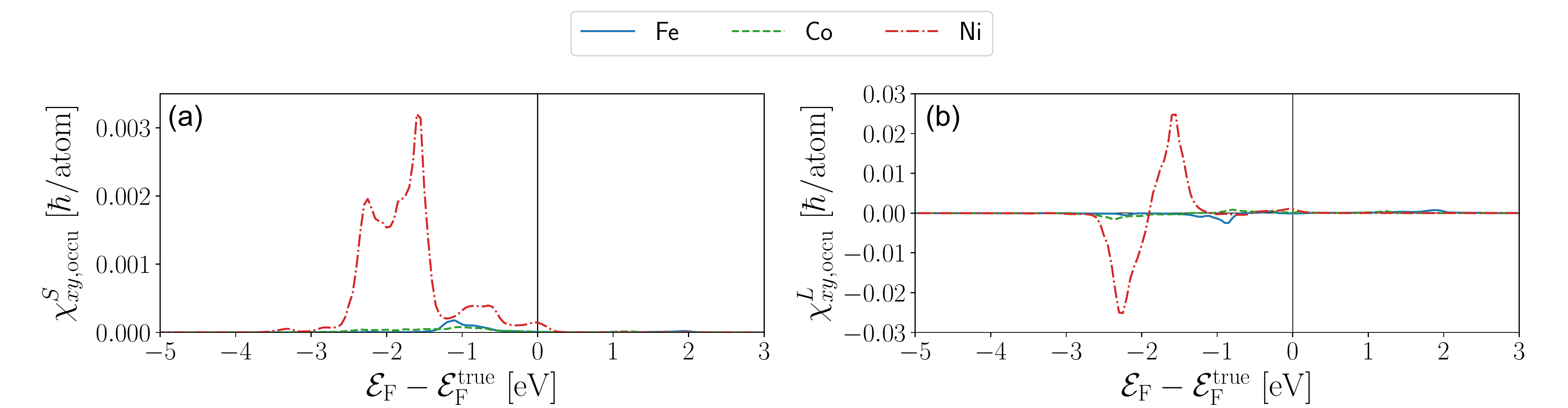}
\caption{
\label{fig:SM_FSB}
The Fermi-energy ($\mathcal{E}_\mathrm{F}$) dependence of the occupation-change contribution [Eq.~\eqref{eq:chi_occupation_final}] for (a) the spin pumping and (b) the orbital pumping in Fe, Co, and Ni. 
}
\vspace{12pt}
\end{figure}

In Fig.~\ref{fig:SM_FSB}, we show the occupation-change contribution in Fe, Co, and Ni, as a function of the Fermi energy. Generally, we find that the values are significantly smaller than the $\mathcal{T}$-even contribution which is subject to the same symmetry constraints. Thus, in Tab. I and Figs. 2 and 3 of the main text, the occupation-change contribution is not included in the result. The occupation-change contribution is small because the occupation-change contribution is driven by the diagonal components of the exchange torque and OAM/spin, which are much smaller than the off-diagonal components. We note that the diagonal component of the exchange torque is zero if the SOC is absent. Thus, even for the spin pumping, the occupation-change contribution requires the SOC, unlike the other contributions. This also explains why the occupation-change contribution is more pronounced in Ni compared to Fe and Co because the ratio of the SOC and the exchange interaction is relatively large in Ni, compared to the ratio in Fe and Co.

Meanwhile, we find that $\chi_{xy,\mathrm{occ}}^S$ is smaller than $\chi_{xy,\mathrm{occ}}^L$ by an order of magnitude, while the state-change contribution exhibits the opposite behavior. This is because a transverse component of the OAM to the magnetization is generally more strongly polarized than that of the spin. In ferromagnets, the spin is substantially split due to the exchange interaction. On the other hand, although the OAM is generally quenched by the crystal field potential, there exist hotspots in which the orbital degeneracy remains protected by the crystal symmetries, in which a transverse component of the OAM may be large~\cite{Go2023long}.

\subsection{Dependence on the spin-orbit coupling}

\begin{figure}[t!]
\vspace{20pt}
\centering
\includegraphics[angle=0, width=0.95\textwidth]{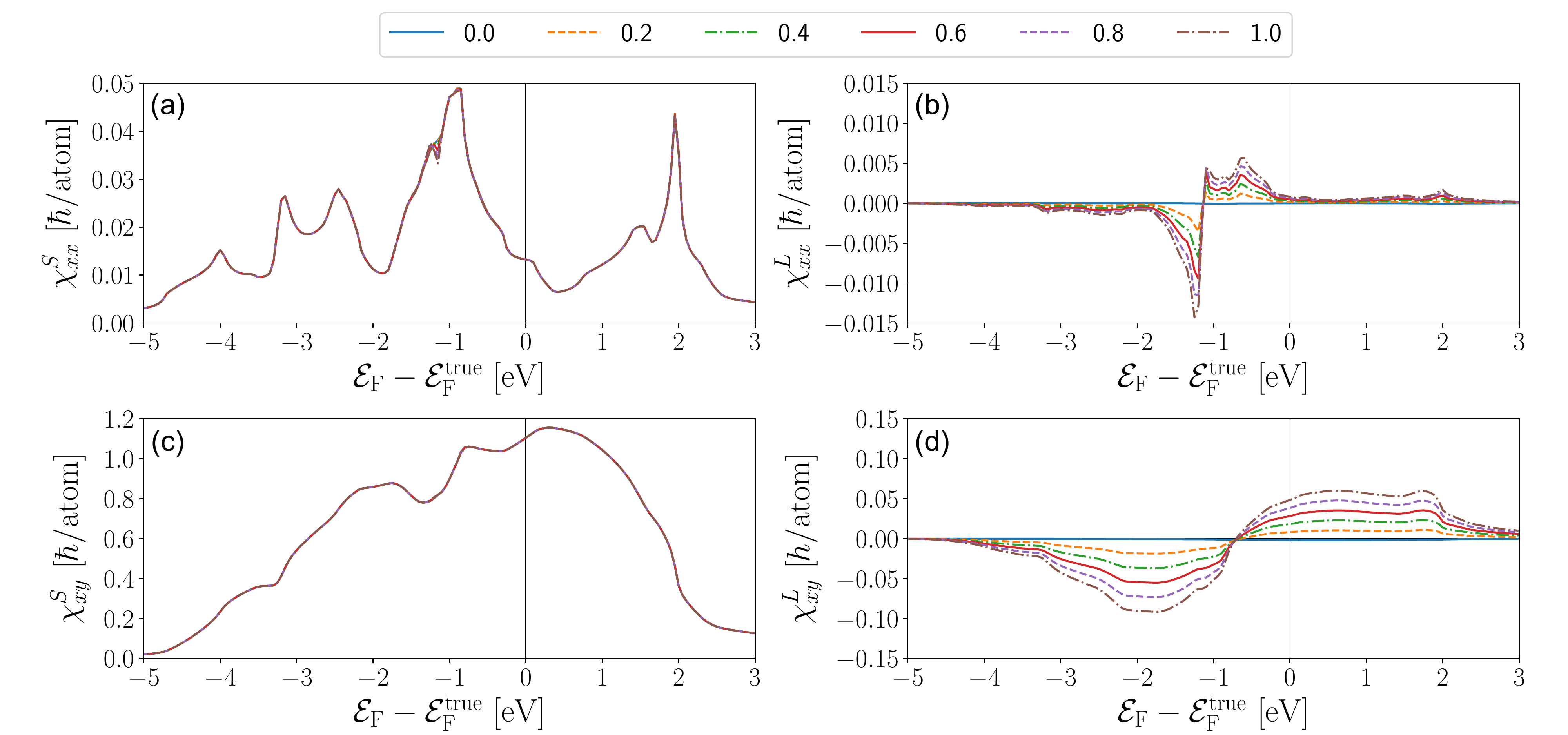}
\caption{
\label{fig:SOC_Fe}
SOC dependence of the spin pumping and orbital pumping in Fe, (a) $\chi_{xx}^S$, (b) $\chi_{xx}^L$, (c) $\chi_{xy}^S$, and (d) $\chi_{xy}^L$. The scaling ratio of the SOC is indicated by different line colors and styles. The Fermi energy ($E_\mathrm{F}$) is varied with respect to the true value $\mathcal{E}_\mathrm{F}^\mathrm{true}$ within the rigid band approximation. The energy broadening is set to $\Gamma=25\ \mathrm{meV}$. 
}
\end{figure}

\begin{figure}[t!]
\vspace{50pt}
\centering
\includegraphics[angle=0, width=0.95\textwidth]{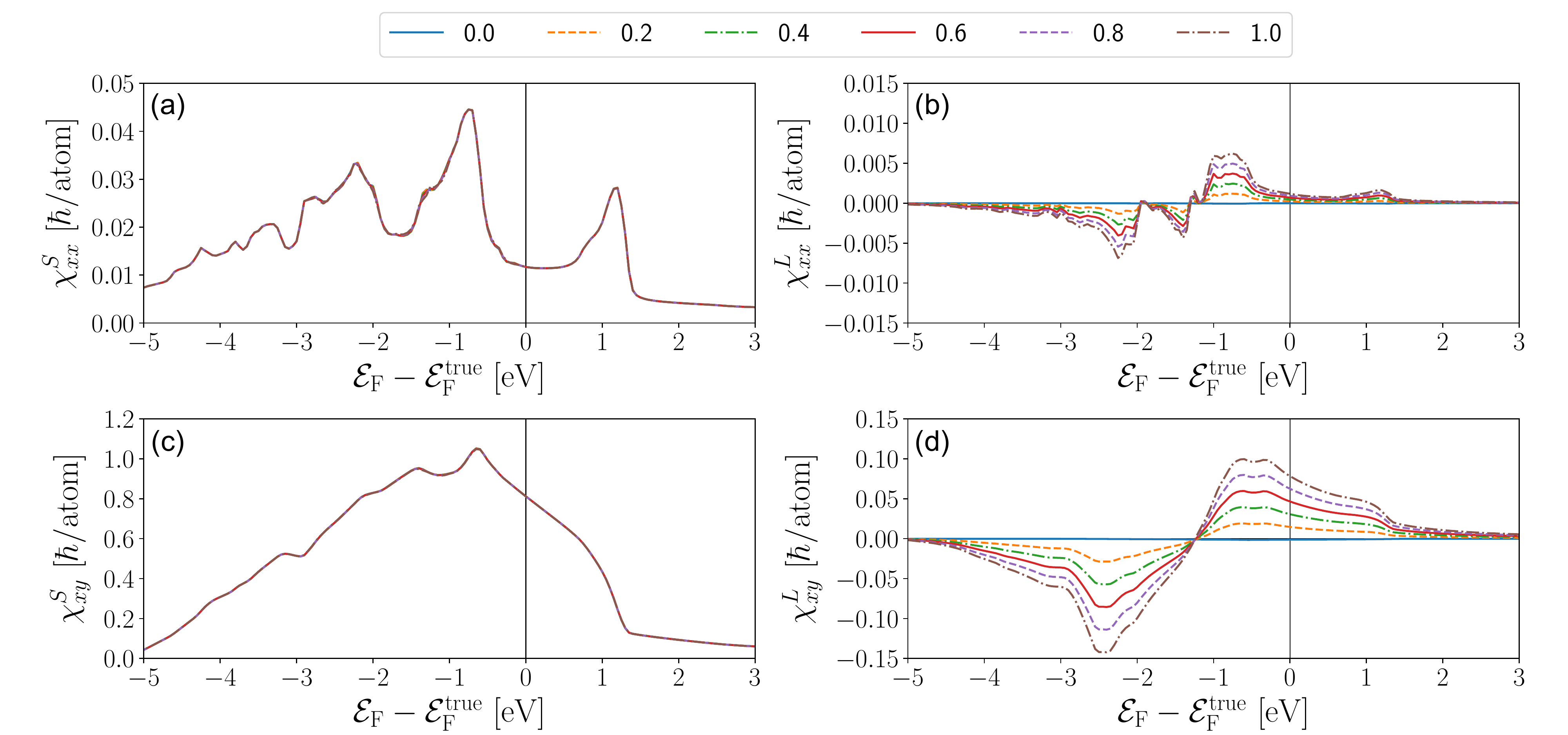}
\caption{
\label{fig:SOC_Co}
SOC dependence of the spin pumping and orbital pumping in Co, (a) $\chi_{xx}^S$, (b) $\chi_{xx}^L$, (c) $\chi_{xy}^S$, and (d) $\chi_{xy}^L$. The scaling ratio of the SOC is indicated by different line colors and styles. The Fermi energy ($E_\mathrm{F}$) is varied with respect to the true value $\mathcal{E}_\mathrm{F}^\mathrm{true}$ within the rigid band approximation. The energy broadening is set to $\Gamma=25\ \mathrm{meV}$. 
}
\end{figure}

\begin{figure}[t!]
\centering
\includegraphics[angle=0, width=0.95\textwidth]{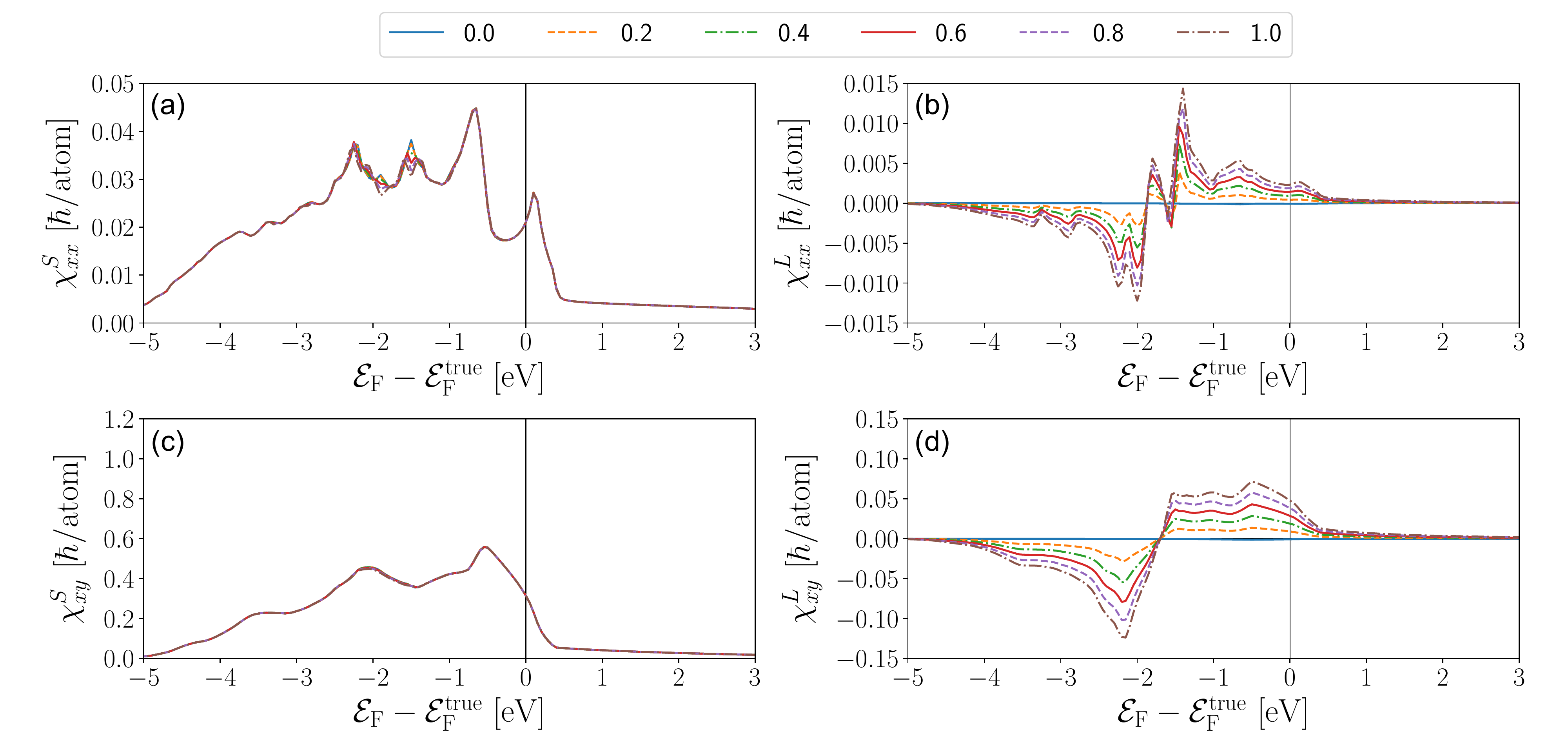}
\caption{
\label{fig:SOC_Ni}
SOC dependence of the spin pumping and orbital pumping in Ni, (a) $\chi_{xx}^S$, (b) $\chi_{xx}^L$, (c) $\chi_{xy}^S$, and (d) $\chi_{xy}^L$. The scaling ratio of the SOC is indicated by different line colors and styles. The Fermi energy ($E_\mathrm{F}$) is varied with respect to the true value $\mathcal{E}_\mathrm{F}^\mathrm{true}$ within the rigid band approximation. The energy broadening is set to $\Gamma=25\ \mathrm{meV}$. 
}
\end{figure}

We compute the $\mathcal{T}$-even and $\mathcal{T}$-odd terms in the state-change contribution by scaling the SOC by 0, 20, 40, 60, 80, and 100~\%. The results for Fe, Co, and Ni are shown in Figs.~\ref{fig:SOC_Fe}, \ref{fig:SOC_Co}, and \ref{fig:SOC_Ni}, respectively. In all cases, we find that the spin pumping is nearly independent of the SOC, while the orbital pumping is linearly proportional to the SOC.

\newpage 

\end{document}